\documentclass[10pt]{emulateapj}
\usepackage{apjfonts}
\usepackage[utf8x]{inputenc}
\usepackage{graphicx}
\usepackage{amssymb}
\usepackage{amsmath}
\usepackage{mathtools}

\newcommand{\lsol}{\textrm{L}_{\odot}}
\newcommand{\msol}{\textrm{M}_{\odot}}

\newcommand{\sfrAll}{\textrm{SFR}_{\textrm{Tot}}\: =\: \textrm{SFR}_{\textrm{NUV}}^{0} + (1-\eta)\textrm{SFR}_{\textrm{IR}}}
\newcommand{\sfrIR}{\textrm{SFR}_{\textrm{IR}}\: = \: 4.6\times 10^{-44}\textrm{}L_{\textrm{IR}} \textrm{ }\msol\textrm{yr}^{-1}}
\newcommand{\sfrUV}{\textrm{SFR}_{\textrm{UV}}^{0}\: = \: 1.2\times 10^{-43}\textrm{}L_{\textrm{NUV, obs}} \textrm{ }\msol\textrm{yr}^{-1}}
\newcommand{\MineoXray}{L_{0.5-8\textrm{keV}}(\textrm{ erg s}^{-1})\: = \: (3.5\pm0.4)\times 10^{39}\textrm{ SFR}(\msol\textrm{yr}^{-1})}
\newcommand{\chisquared}{\ensuremath{\chi^2}} 

\begin{document}
\title{XMM-Newton Observations of Three Interacting Luminous Infrared Galaxies}
\author{{Dale Mudd}\altaffilmark{1}, {Smita Mathur}\altaffilmark{1,2}, {Matteo Guainazzi}\altaffilmark{3}, {Enrico Piconcelli}\altaffilmark{4}, {Stefano Bianchi}\altaffilmark{5}, {S. Komossa}\altaffilmark{6}, {Cristian Vignali}\altaffilmark{7}, {Giorgio Lanzuisi}\altaffilmark{8}, {Fabrizio Nicastro}\altaffilmark{4,9}, {Fabrizio Fiore}\altaffilmark{10}, {Roberto Maiolino}\altaffilmark{11,12}}

\email{mudd@astronomy.ohio-state.edu}

\altaffiltext{1}{Department of Astronomy, The Ohio State University, Columbus, Ohio 43210, USA}
\altaffiltext{2}{Center for Cosmology and Astro-Particle Physics, The Ohio State University, Columbus, OH 43210}
\altaffiltext{3}{European Space Astronomy Centre of ESA, PO Box 78, Villanueva de la Ca\~{n}ada, 28691 Madrid, Spain}
\altaffiltext{4}{Osservatorio Astronomico di Roma - INAF, Via di Frascati 33, 00040, Monte Porzio Catone, RM, Italy}
\altaffiltext{5}{Dipartimento di Fisica, Universit\`{a} degli Studi Roma Tre, via della Vasca Navale 84, 00416 Roma, Italy}
\altaffiltext{6}{Max-Planck Institut f\"{u}r Radioastronomie, Auf dem H\"{u}gel 69, D-53121 Bonn, Germany}
\altaffiltext{7}{INAF – Osservatorio Astronomico di Bologna, Via Ranzani 1, 40127 Bologna, Italy}
\altaffiltext{8}{INAF – Istituto di Astrofisica Spaziale e Fisica Cosmica di Bologna, via Gobetti 101, 40129 Bologna, Italy}
\altaffiltext{9}{Harvard-Smithsonian Center for Astrophysics, 60 Garden St., MS-04, Cambridge, MA 02138, USA}
\altaffiltext{10}{INAF – Osservatorio Astronomico di Roma, Via di Frascati 33, 00040 Monteporzio Catone, Italy}
\altaffiltext{11}{Department of Physics, University of Cambridge, 19 JJ Thomson Avenue, Cambridge, CB3 0HE, United Kingdom}
\altaffiltext{12}{Kavli Institute for Cosmology, Madingley Road, Cambridge, CB3 0HA, United Kingdom}

\date{\today}

\begin{abstract}
We investigate the X-ray properties of three
interacting luminous infrared galaxy systems.  In one of these systems,
IRAS 18329+5950, we resolve two separate sources.  A second and third source, IRAS
19354+4559 and IRAS 20550+1656, have only a single X-ray source
detected.  We compare the observed emission to PSF profiles and
determine that they are all consistent with the PSF, 
albeit with large uncertainties for some of our sources.  
We then model the spectra to determine soft (0.5--2
keV) and hard (2--10 keV) luminosities for the resolved sources, and
compare these to relationships found in the literature between infrared
and X-ray luminosities for starburst galaxies.  We obtain
luminosities (0.5--10 keV) ranging from $1.7-7.3\times10^{41}$ 
erg/s for our systems.  These
X-ray luminosities are consistent with predictions for
star-formation-dominated sources and thus are most likely due to starbursts, 
but we cannot conclusively rule out AGN.
\end{abstract}


\section{Introduction}
\label{sec: intro}
Over the past few decades, much work has emphasized the role and importance of mergers in 
shaping galactic evolution (e.g., \citealp{Toomre72, Ellison11, Hopkins06}).  In particular, mergers have been presented as potential drivers of 
starbursts and active galactic nuclei (AGN; \citealp{Bauer04, Summers04, Bell12, Komossa03, Teng09, Piconcelli10}).  
Broadly, hierarchical growth simulations such as those in \citet{Hopkins06} paint a picture in which
galactic mergers provide gas inflows to ignite starbursts, which are then followed by growth of the central supermassive black holes (SMBHs)
of galaxies, sometimes leading to the birth of an AGN, until some feedback mechanism, whose origin and existence is contentious, possibly stops the process.  
The interactions of two galaxies can also cause their central SMBHs to become active simultaneously, resulting in a dual
AGN \citep{Komossa06}.  A complication in this scenario is that some simulations suggest that most of the dual AGN activity 
is non-simultaneous, particularly during their highest luminosity phases when they would be easiest to
detect observationally \citep{VanWassenhove12}.  Other simulations 
have hinted that this merger scenario was more important in the past, when interacting systems were much more gas-rich, but that 
it still plays a role in the local universe for lower-luminosity AGN \citep{Draper12}.  

Luminous infrared galaxies (LIRGs), galaxies that have infrared luminosities $L_{IR}$ $\geq$ $10^{11}$ $\lsol$ (for a review, see 
\citealp{Sanders96} and references therein), where IR ranges here span 8-1000 $\mu$m, are often found to be mergers and have 
distinctive morphologies \citep{Yuan10}.  The emission from such systems is prominent in the infrared (IR) portion of the 
spectrum due to heavy dust reprocessing.  The Great Observatories All-sky LIRG Survey (GOALS), which 
has been observing LIRGs across the electromagnetic spectrum (e.g., \citealp{Armus09} and \citealp{Iwasawa11}), and other studies (e.g., 
\citealp{Ptak03, Braito09, AlonsoHerrero12, Charmandaris10}) have attempted to determine if 
the drivers of the high IR luminosities in these systems are central starbursts or AGN.  In one such system, NGC 6240, two separate AGN 
were detected by \citet{Komossa03} at a separation of about 1.4 kpc.  This discovery has fueled much discussion about the nature of dual 
AGN, their prevalence, and lifetimes.  
  
Several studies have investigated the connection between AGN activity and environment.  When looking at quasars, 
\citet{Serber06} found a local overdensity of galaxies of $\sim$3 around the brightest objects and 
an overdensity of $\sim$1.4 for fainter quasars compared to an average galaxy. They show that the overdensity of galaxies around quasars is the same as the 
overdensity around L* galaxies on scales of $\sim$1 Mpc.  At smaller radial separations, however, quasars have a larger overdensity of nearby galaxies than L* galaxies 
do at the same separations.  \citet{Ellison11} reported a complementary result focusing on AGN; they found AGN are 2.5 times more likely to 
be in pairs than similar but inactive galaxies.  Using Swift's Burst Alert Telescope (BAT) AGN sample, \citet{Koss11} 
also found evidence that strong AGN activity is associated with mergers and interactions.  Specifically, 
they report that 24\% of BAT AGN are undergoing a merger, whereas this number is closer to 1\% for a sample of normal galaxies.  
Similarly, looking at a subsample from SDSS DR7, \citet{Liu12} 
found that young stellar ages, star formation, and SMBH activity in a given galaxy correlate well with smaller separation from its 
nearby neighbors.  These and other works (e.g., \citealp{Ajello12}) suggest that a galaxy's environment can play a non-negligible role in 
fueling AGN, whether through interactions or mergers with nearby galaxies.           

Because of heavy obscuration at most wavelengths in dust-enshrouded systems, X-rays, which are less attenuated, 
become a useful tool to probe behind optically thick screens.  From a sample of the optically 
luminous quasars from SDSS DR3, spanning a redshift of 1.5-4.5, the average power law index for X-ray emission was found to 
be $1.9^{+0.1}_{-0.1}$ \citep{Just07}.  The spread in the 2--10 keV photon index, however, is quite large, with observed 
systems running the range of $\Gamma$ $\sim$ $1.5-2.5$.  A proposed explanation for this 
spectral behavior is that an accretion disk feeding the central SMBH thermally emits in the 
optical and UV.  These photons then inverse Compton upscatter to X-rays off of a hot plasma surrounding the disk 
\citep{Shapiro76, Sunyaev80, Haardt93}.  The rather large spread in the value of photon index 
is thought to arise from the specific accretion rate of the SMBH regulating disk cooling-- a high accretion rate increases 
the rate of disk emission, providing more soft (here, 0.5--2 keV) photons and increasing the Compton cooling of the corona.  This, then, reduces the 
abundance of hard photons.  Combined with the increase of soft photons, higher accretion rates are thus associated with a steeper index 
\citep{Williams04, Shemmer08}.  Much of a galaxy's SMBH growth is believed to occur in regions heavily obscured by dust \citep{Hopkins06, Fabian99}, 
so searching in X-rays is one of the few viable options for spotting this enshrouded phase, provided the column density is not overly large 
($\gtrsim 10^{24}\: \textrm{cm}^{-2}$).

AGN activity is not the only interesting phenomenon strongly associated with galactic interactions.  
Another prominent event often triggered is a marked increase in the star formation rate of one or 
both component galaxies, provided neither is gas poor.  If this increase is large enough, the galaxy becomes a starburst.  
\citet{Muzzin12} suggest that, while the gross properties of star-forming galaxies are tied to their stellar 
masses, their environment appears to regulate the fraction of systems that are starbursting.  

Like AGN, star-forming regions also produce specific X-ray signatures (e.g., \citealp{Persic02}).  The ionizing photons from supernovae 
and short-lived O and B stars form a thermal plasma typically with temperature between 0.1-1 keV.  A secondary power law component, associated with 
high mass X-ray binaries (HMXBs), indicative of recent star formation \citep{Fabbiano06}, is often observed in starburst systems as well.  These
are said to trace recent or ongoing star formation because they require an OB star around a neutron star (NS) or black hole (BH) and thus only have
a lifetime of $10^{6-7}$ years.  HMXBs have been observed with a range of photon indices, typically lying within $\Gamma\textrm{ = }$ 1-2 with the steepest being 
greater than 2.4 \citep{Remillard06}.  Low mass X-ray binaries (LMXBs), a possible contaminant, are usually described by either a power law with index $\sim$1.6 
or bremsstrahlung at 7.3 keV \citep{Irwin03, Fabbiano06, Persic03}.  These LMXB spectral indices can run as steep as 2 in the highest luminosity cases.  LMXBs do not trace recent 
star formation, however, since they are connected to lower mass companions around a NS or BH and thus have lifetimes that are 100--1000 times longer than their 
higher mass counterparts.   

\citet{Kennicutt98} associated star forming regions with IR emission due to dust reprocessing of UV and optical emission from young stars.  Several studies (e.g, \citealp{Ranalli03}) 
have proposed X-ray/IR luminosity relationships in IR-bright galaxies, arguing that star formation is traced in the IR from reprocessing and 
in X-rays from HMXBs, provided there is no AGN present.  The use of X-rays as a tracer of star formation also requires that LMXBs are a negligible contributor, 
which is generally a safe assumption for the high star formation rates encountered in starbursting systems.  One can then use these 
relations to predict the X-ray output of a system originating from star formation alone.  It should be noted, however, 
that the existence of either a starburst or an AGN does not discount the other \citep{Sani10, Santini12}.  Additionally, some argue for a more complete 
picture taking into account escaping, non-reprocessed UV radiation when attempting to synthesize star formation data for certain 
luminosity regimes (e.g., \citealp{Vattakunnel12, Mineo12I}).

In this paper we use the association between AGN activity, starbursts, mergers, and, 
by extension, large IR luminosity, to search for dual AGN.  We also investigate how well our data are fit by 
a starbursting population alone.  In  \textsection\ref{sec: samp}, we describe our sample
selection and discuss the systems we examined in detail.  \textsection\ref{sec: reduction} presents image and spectral analysis as well as 
the luminosities we find.   \textsection\ref{sec: disc} compares our data to existing IR/X-ray relations and discusses our prospectives
on dual AGN detection.  Finally, we conclude in \textsection\ref{sec: conclusion}.

We adopt a cosmology of $H_{0}$ $= 70\textrm{ km s}^{-1}\textrm{ Mpc}^{-1}$, $\Omega_{\textrm{M}}$ $= 0.27$, and $\Omega_{\Lambda}$ $= 0.73$ and 
all relationships taken from the literature assume a Salpeter IMF.

\section{Sample}
\label{sec: samp}
We derived our sample from a subset of relatively local IR bright galaxies presented in \citet{Arribas04}.  Our systems 
are nearby (redshifts of 0.03 < z < 0.07), to ensure a high number of photon counts, and are known to be interacting with 
two optical nuclear regions, though there is some uncertainty about the location of the secondary nucleus in IRAS 20550+1656.  All are LIRGs,
and many have roughly measured nuclear separations. We selected based on interaction class using the scheme proposed in \citet{Surace98} and 
refined in \citet{Veilleux02}, restricting our sample to class IIIa and IIIb systems with nuclear separations around 10 kpc.  A class IIIa system 
is composed of two or more galaxies with evidence of tidal interactions (tails, bridges) and with nuclear separation of greater than 10 kpc, whereas a class 
IIIb system has similar tidal structures but with nuclear separation less than 10 kpc. Additionally, we required that systems have angular separations 
that would likely be resolvable with \emph{XMM-Newton} ($\sim8-10^{\prime\prime}$).  After this, we selected the LIRGs with the highest
IR fluxes, and, by association, likely the most X-ray counts barring the system being Compton thick.  The IR luminosity for all the potential targets was calculated 
from the IRAS Revised Bright Galaxy Sample (hereafter RBGS; \citealp{Sanders03}) using the distance and IR flux defined in the RBGS as
\begin{equation}
\begin{split}
 F_{\textrm{IR}}\: = \: &1.8 \times 10^{-11}\left(13.48\frac{f_{12\mu\textrm{m}}}{\textrm{Jy}} \:+\: 5.16\frac{f_{25\mu\textrm{m}}}{\textrm{Jy}} \:+\:  \right. \\
                   & \left. 2.58\frac{f_{60\mu\textrm{m}}}{\textrm{Jy}} \:+\: \frac{f_{100\mu\textrm{m}}}{\textrm{Jy}}\right) \textrm{ erg cm}^{-2}\textrm{s}^{-1}.
\end{split}
\label{eq: IRflux}
\end{equation}

Three systems met our criteria: IRAS 18329+5950, IRAS 19354+4559, and IRAS 20550+1656.  We observed these with \emph{XMM-Newton} and restricted all 
analyses to 0.5--10 keV so as to stay in range of the detector's optimal sensitivity.  Table \ref{tab: ObservationData} summarizes our data, including the 
observation ID, exposure times, nuclear separations, and counts for each of our sources.

IRAS 18329+5950 is a class IIIa system composed of two merging galaxies with estimated nuclear separations of $28^{\prime\prime}$ \citep{Arribas04}.  
The offset in declination ($5.^{\!\!\prime\prime}70$) of the two galaxies in this system is much smaller than the offset in right ascension ($27.^{\!\!\prime\prime}5$).  In projection, 
it appears that the nucleus of the eastern component (18329E) is close to the plane of the disc of the western 
component (18329W), as can be seen in the bottom row of Figure \ref{fig: Images}. The system is classified as a LIRG in the RBGS. Using the RBGS fluxes, we 
calculate an IR luminosity of $\log(L_{\textrm{IR}}/\lsol)$ $= 11.60$ based on the redshift of z=0.029, 
which corresponds to a distance of $\sim$129 Mpc \citep{Armus09}.  Optical line ratios, discussed further in \textsection\ref{subsec: resumen}, 
suggest that this system is a starburst or starburst/AGN composite.  

IRAS 19354+4559 is a class IIIb merger system with nuclear separation of approximately $8.^{\!\!\prime\prime}5$ \citep{Arribas04}.  
In this case, the offsets in right ascension and declination are comparable, being $6.^{\!\!\prime\prime}7$ and $5.^{\!\!\prime\prime}3$, respectively.  
Both galaxies have significant tidal features in the optical, suggestive of mutual interaction.  They appear as mostly edge-on disks, 
although the eastern component has several foreground stars contaminating the optical images, partially confusing the orientation, which can be 
seen in Figure 1 of \citet{Arribas04}.  Using fluxes from the RBGS, we calculate an upper limit IR luminosity of $\log(L_{\textrm{IR}}/\lsol)$ $= 11.85$ based on the redshift of z=0.067, 
which corresponds to a distance of $\sim$289 Mpc \citep{Lawrence99}.  We did not find any IR or optical line diagnostics for this system.

Similarly, IRAS 20550+1656 is also a class IIIb merger system.  The separation 
between the optical nuclei is uncertain, but believed to be about $11.^{\!\!\prime\prime}7$ \citep{Arribas04}.  From the RBGS fluxes, we calculate 
the IR luminosity to be $\log(L_{\textrm{IR}}/\lsol) = 11.94$ for a redshift of z=0.036, corresponding to a distance of $\sim$161 Mpc \citep{Armus09}.  
Optical \citep{Baan98} and IR \citep{Inami10} spectroscopy and line ratio diagnostics indicate that this system is a starburst.  The Spitzer data in 
particular \citep{Inami10} suggests that 80\% of the IR luminosity comes from a region outside the nuclei of the two interacting galaxies in this merger.       

\begin{deluxetable*}{lrrcrrrrrr}
\tabletypesize{\scriptsize}
\tablewidth{0pt}
\tablecolumns{9}
\tablecaption{Observational Data}
\tablehead{
\colhead{IRAS Name} &
\colhead{Observation ID} &
\colhead{Date} &
\colhead{Exposure} &
\colhead{Distance} &
\colhead{Separation} &
\colhead{Separation} &
\colhead{Hardness Ratio} &
\colhead{Source Counts} \\
\colhead{} &
\colhead{(XMM)} &
\colhead{} &
\colhead{(ks)} &
\colhead{(Mpc)} &
\colhead{} &
\colhead{(kpc)} &
\colhead{} &
\colhead{(0.5-10 keV)} 
}
  
\startdata

18329+5950E & 0670140301 & 2011-05-04 & 5.3  & 129   & $28^{\prime\prime}$       & 17.4  & $-0.80^{+0.17}_{-0.18}$ & $99\pm13$   \\ 
18329+5950W & 0670140301 & 2011-05-04 & 5.3  & 129   & $28^{\prime\prime}$       & 17.4  & $-0.83^{+0.17}_{-0.17}$ & $106\pm14$  \\
19354+4559  & 0670140501 & 2011-05-10 & 15.2 & 289   & $8.^{\!\!\prime\prime}5$  & 11.9  & $-0.87^{+0.14}_{-0.13}$ & $46\pm12$   \\
20550+1656  & 0670140101 & 2011-10-28 & 60.5 & 161   & $12^{\!\!\prime\prime}$ & 9.13  & $-0.46^{+0.02}_{-0.02}$ & $3461\pm62$ \\

\enddata

\tablecomments{Separation refers to the projected nuclear separation between the interacting 
galaxies; the first is angular separation while the second is physical separation.  The hardness ratio 
represents the overall contribution of hard X-ray emission (2-10 keV) compared to soft (0.5-2 keV) 
emission.  Note that exposure times are good time intervals, explained in the text, and that the source counts and hardness ratios are 
taken from background-subtracted data.}
\label{tab: ObservationData}
\end{deluxetable*}

\section{\emph{XMM-Newton} Observations}
\label{sec: reduction}

\subsection{Image Analysis}
\label{sec: imanal}
\begin{figure*}
  \centerline{
    \includegraphics[width = 5.5cm]{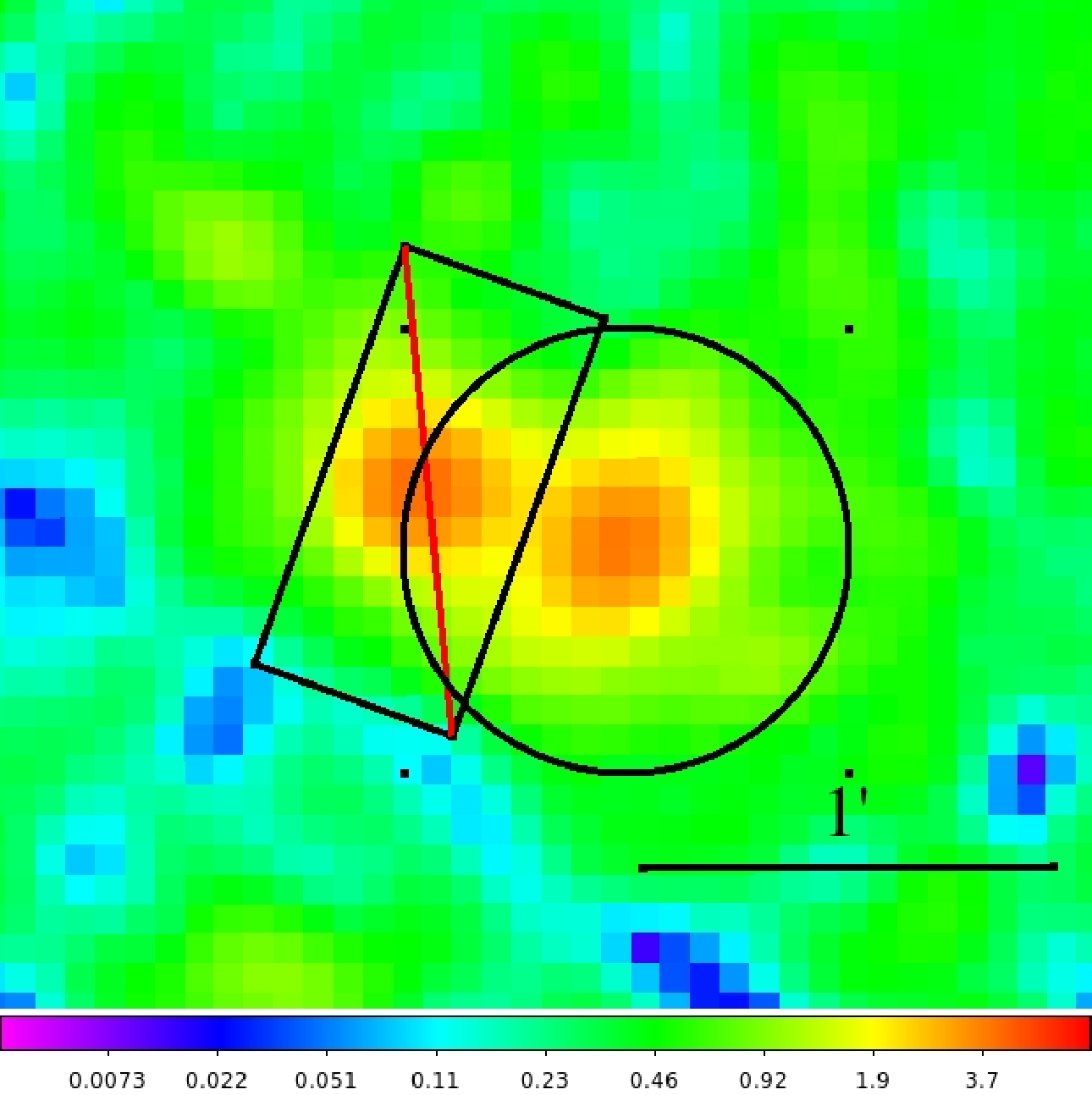}
    \includegraphics[width = 5.5cm]{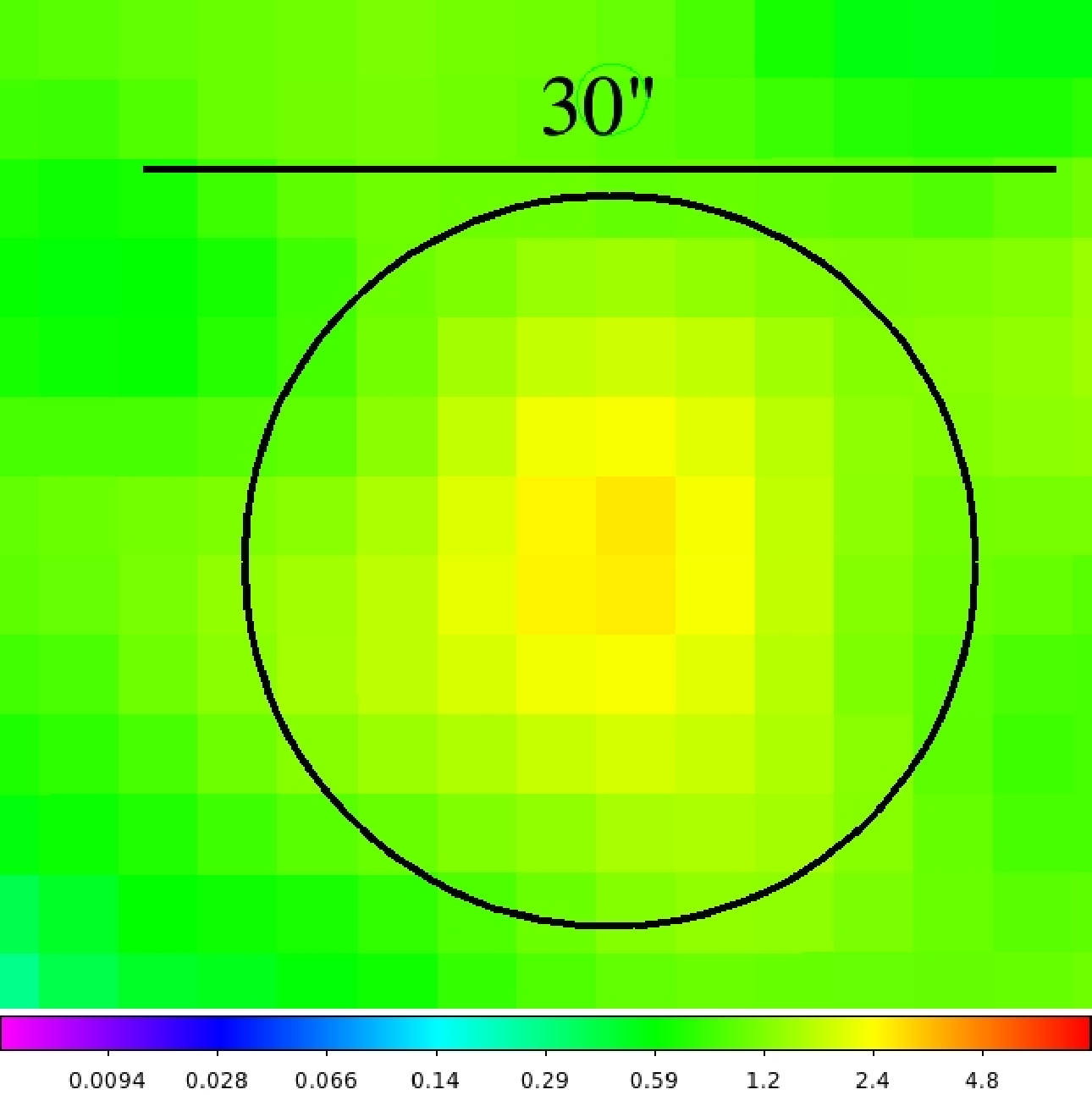}
    \includegraphics[width = 5.5cm]{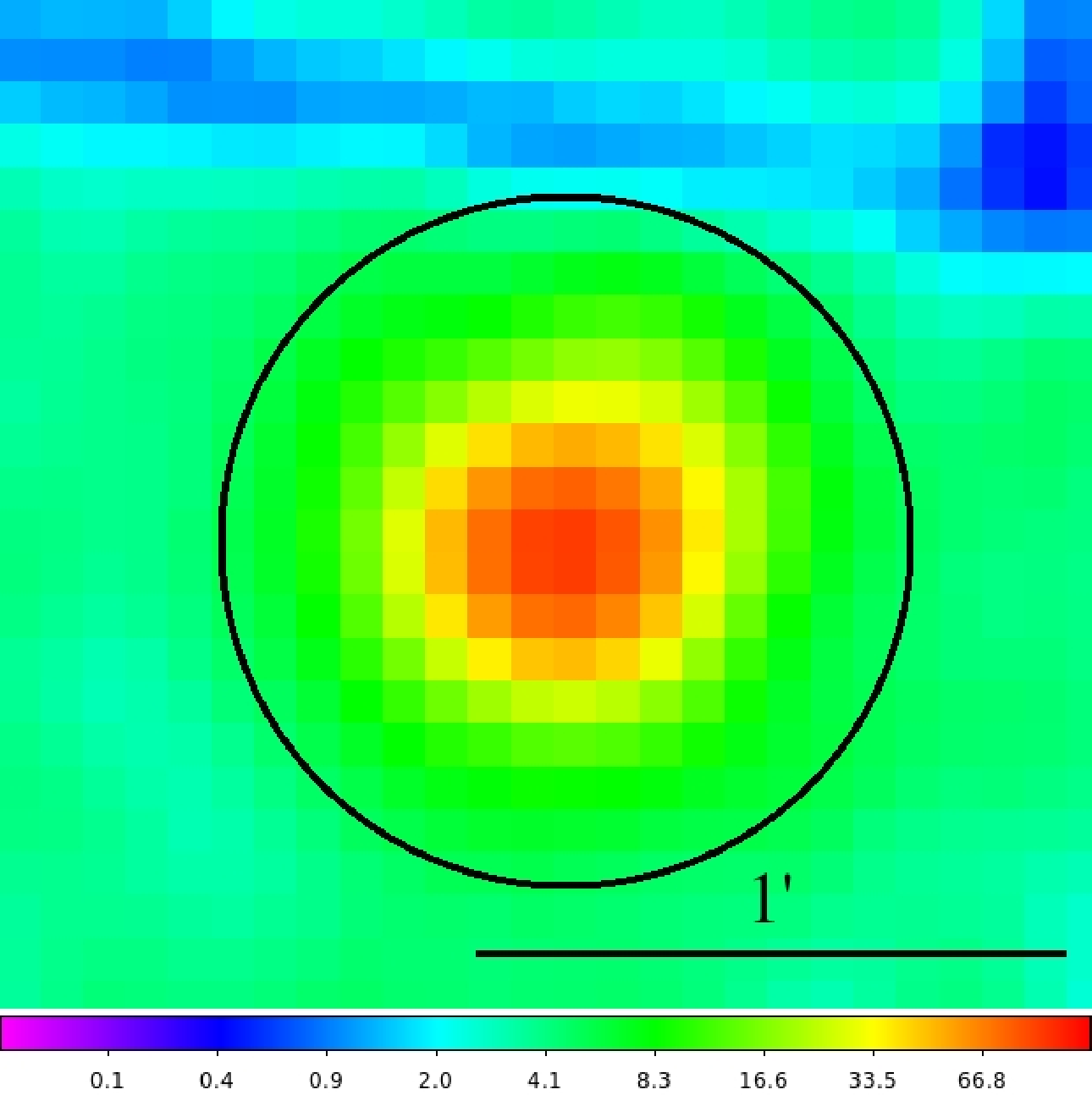}
  }
  \centerline{
    \includegraphics[width = 5.5cm]{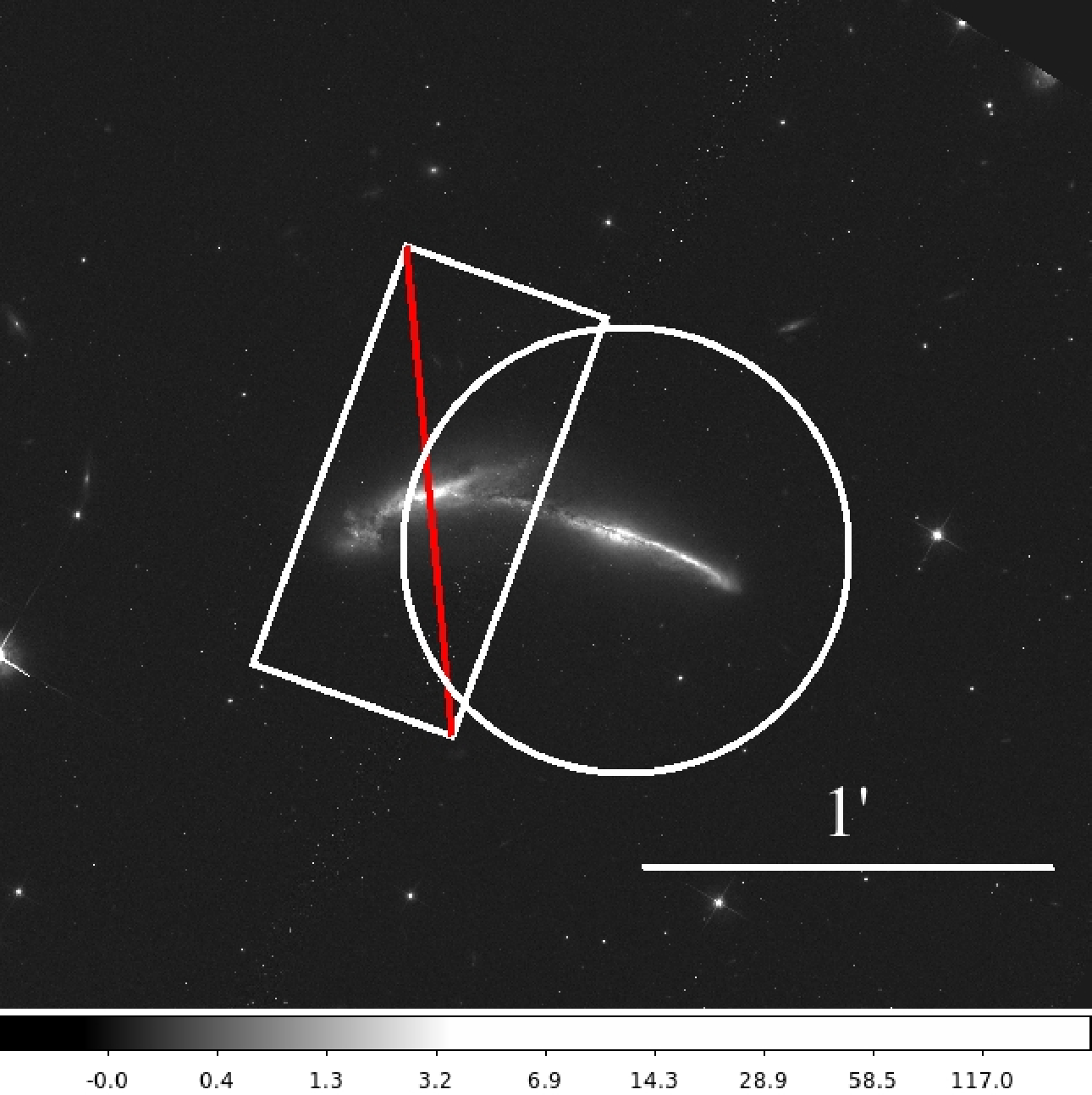}
    \includegraphics[width = 5.5cm]{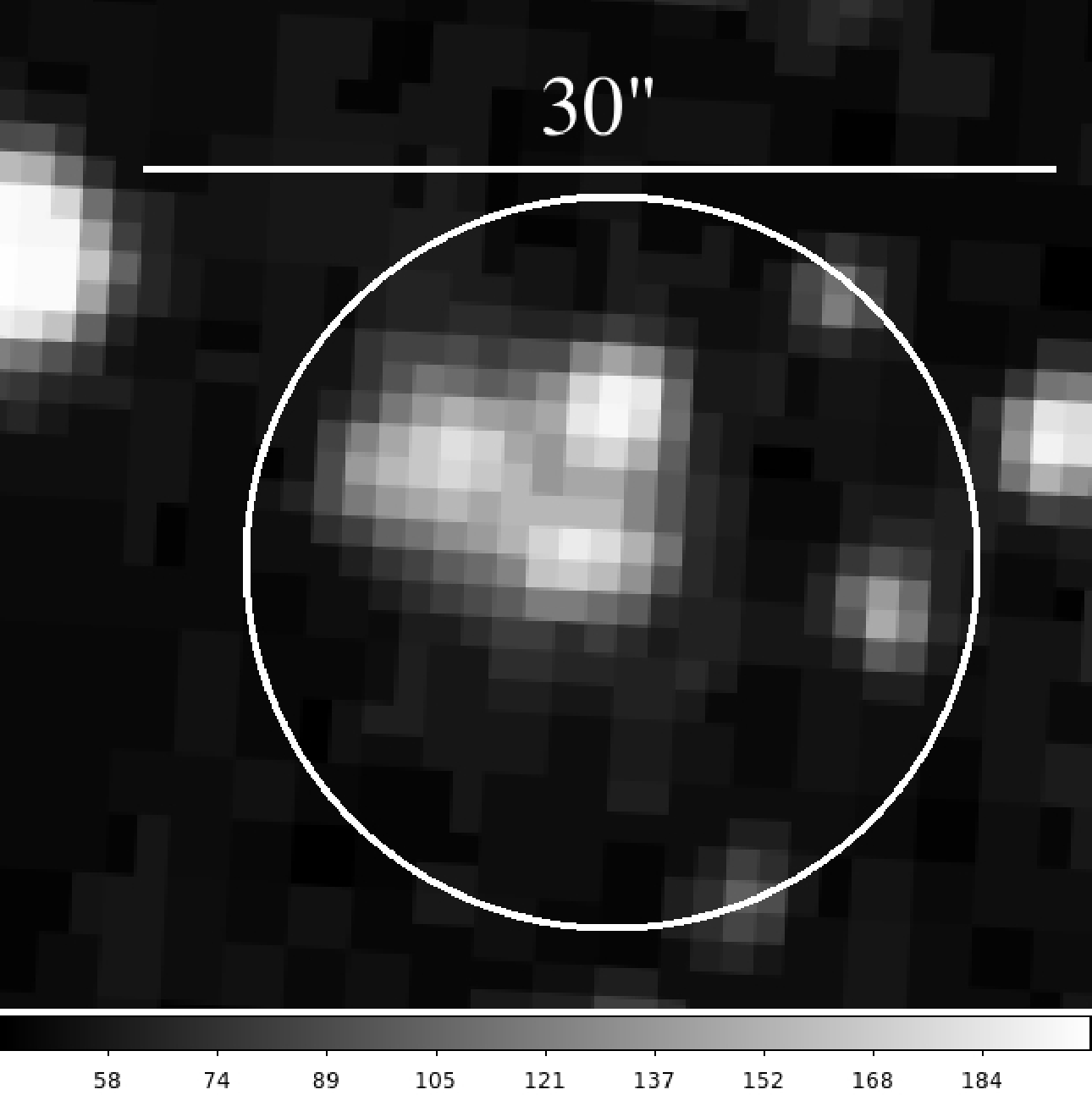}
    \includegraphics[width = 5.5cm]{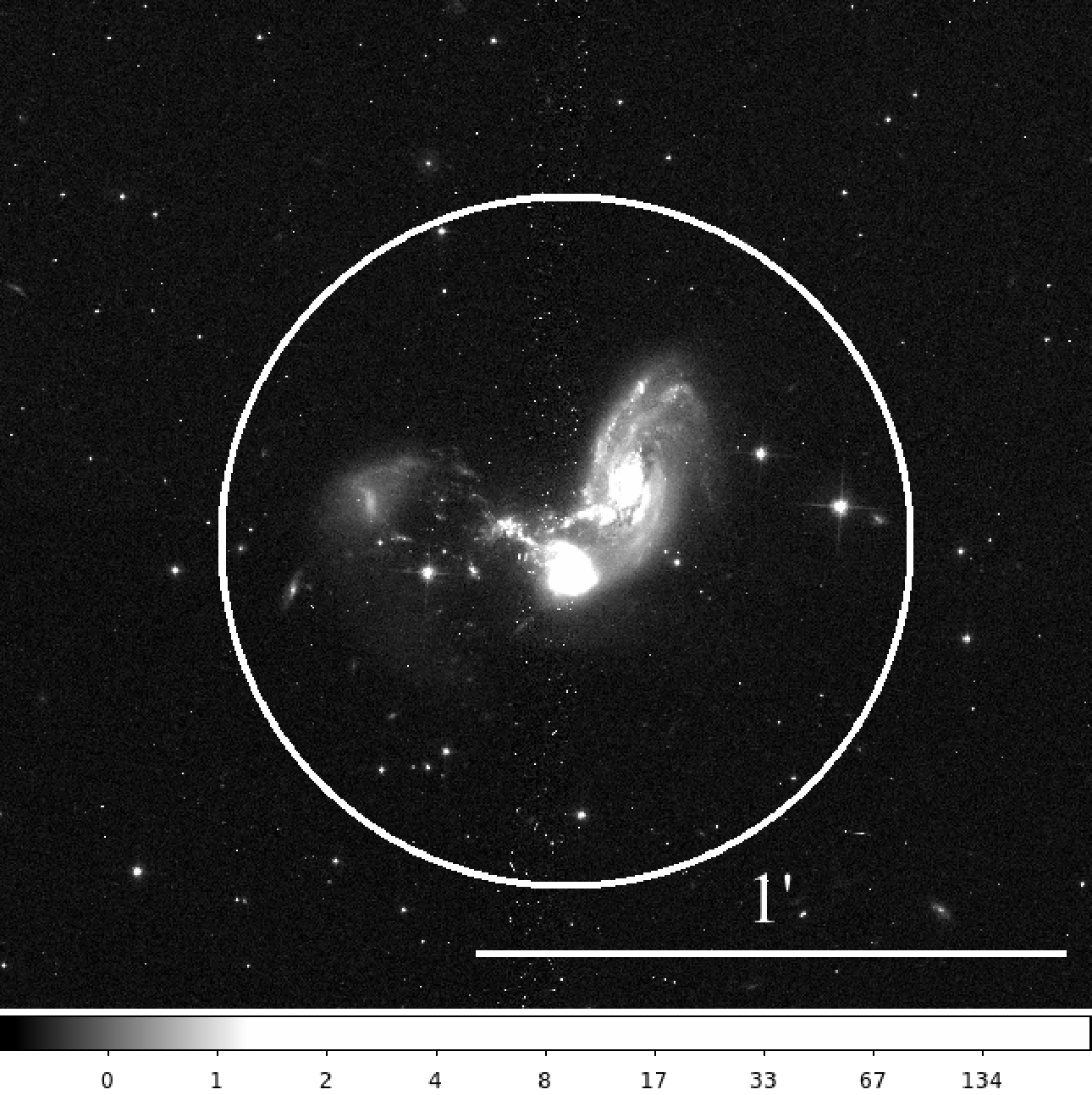}
  }
  \caption{\emph{Top}: Smoothed, false-color \emph{XMM-Newton} images of the three objects in our sample with the spectral extraction region, described in \textsection\ref{sec: reduction}, 
overplotted as black circles.  All images are scaled logarithmically over the range 0.5--10 keV and smoothed using a Gaussian with a kernel size of three pixels.  The units on the 
scale at the bottom of each image is counts, and the physical scale on each image is as labeled.  \emph{Left}: IRAS 18329+5950  \emph{Middle}: IRAS 19354+4559  \emph{Right}: IRAS 20550+1656.  
\emph{Bottom}: Optical images of the galaxies presented in the top row.  IRAS 18329+5950 (\emph{Left}) and IRAS 20550+1656 (\emph{Right}) were taken from archival Hubble Space Telescope images and show 814W-435W filters.  There is no image for IRAS 19354+4559 (\emph{Middle}) in the Hubble archive, so this optical image was taken from the Digitized Sky Survey (DSS).  Note here that the units on the scales are count rates rather than counts for the HST images.  For a better look at the tidal features of this system in particular, we refer the reader to \citet{Arribas04}.}
  \label{fig: Images}
\end{figure*}

All data were taken with the \emph{EPIC} instruments on board \emph{XMM-Newton} and reduced using XMM SAS Version 12.0.1.  
As is typical in X-ray image analysis, we filtered the event files to discard any time interval with $\gtrsim0.5\textrm{ counts s}^{-1}$ in
order to minimize contamination from cosmic rays and other short, time-variant noise sources.  As the \emph{EPIC MOS} cameras have a smaller pixel scale 
than the \emph{EPIC PN} by nearly a factor of four, we first tried creating images and radial profiles using \emph{EPIC MOS}.  Unfortunately, 
there were not enough counts in the \emph{MOS} images to robustly determine the radial profiles, so we proceeded to use the more abundant data from the \emph{PN} detector 
for our analysis.  Processed \emph{EPIC PN} images in the entire 0.5--10 keV band for the three sources in
our sample are shown in Figure \ref{fig: Images}.  The top-left panel of Figure \ref{fig: Images} demonstrates that the nuclear 
regions of the two separate galaxies in IRAS 18329+5950 are resolved in X-rays.  Additionally, the top-middle and top-right panels of Figure \ref{fig: Images} 
show that for both IRAS 19354+4559 and IRAS 20550+1656, the nuclei from the individual component galaxies are not distinctly separable.  
Thus, only a single area could be analyzed for each of these sources.

\begin{figure*}
  \centerline{
    \includegraphics[width = 8.0cm]{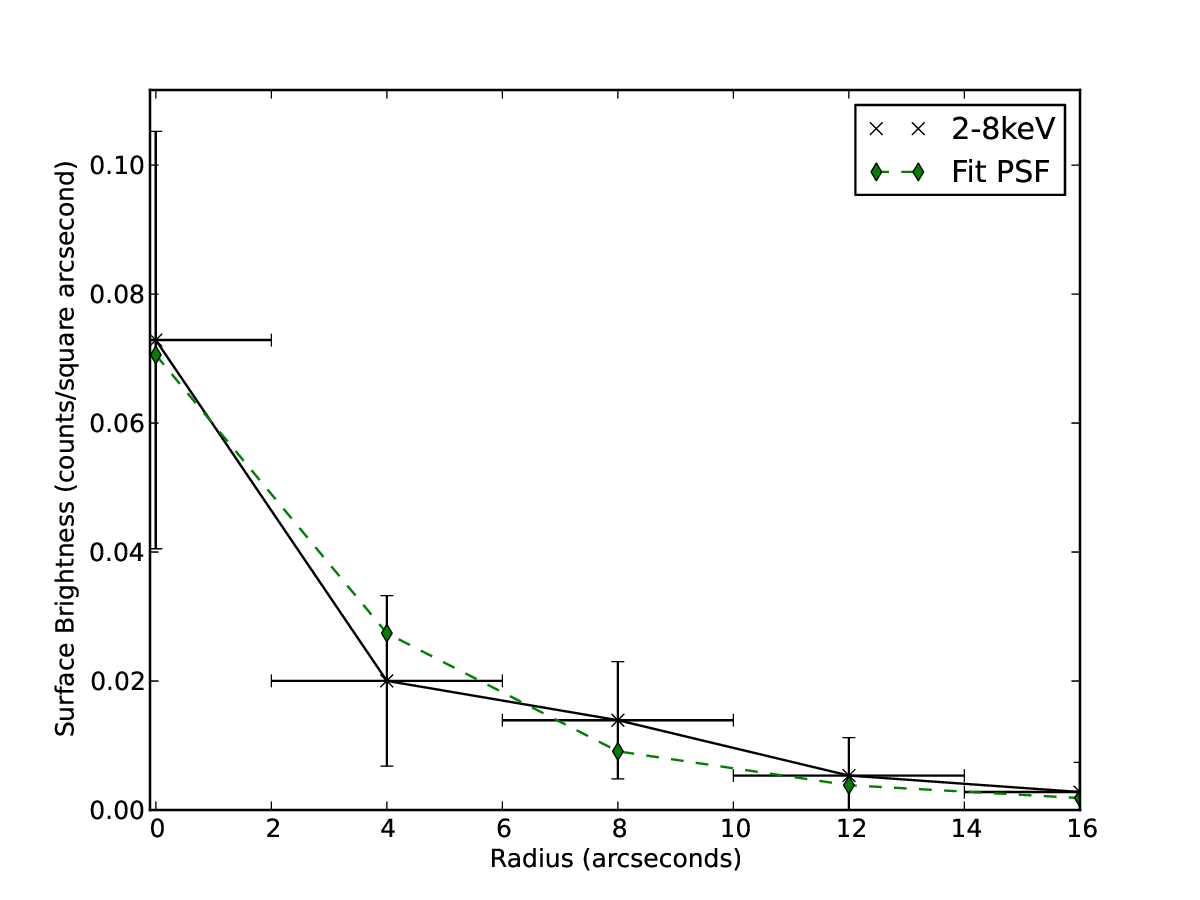}
    \includegraphics[width = 8.0cm]{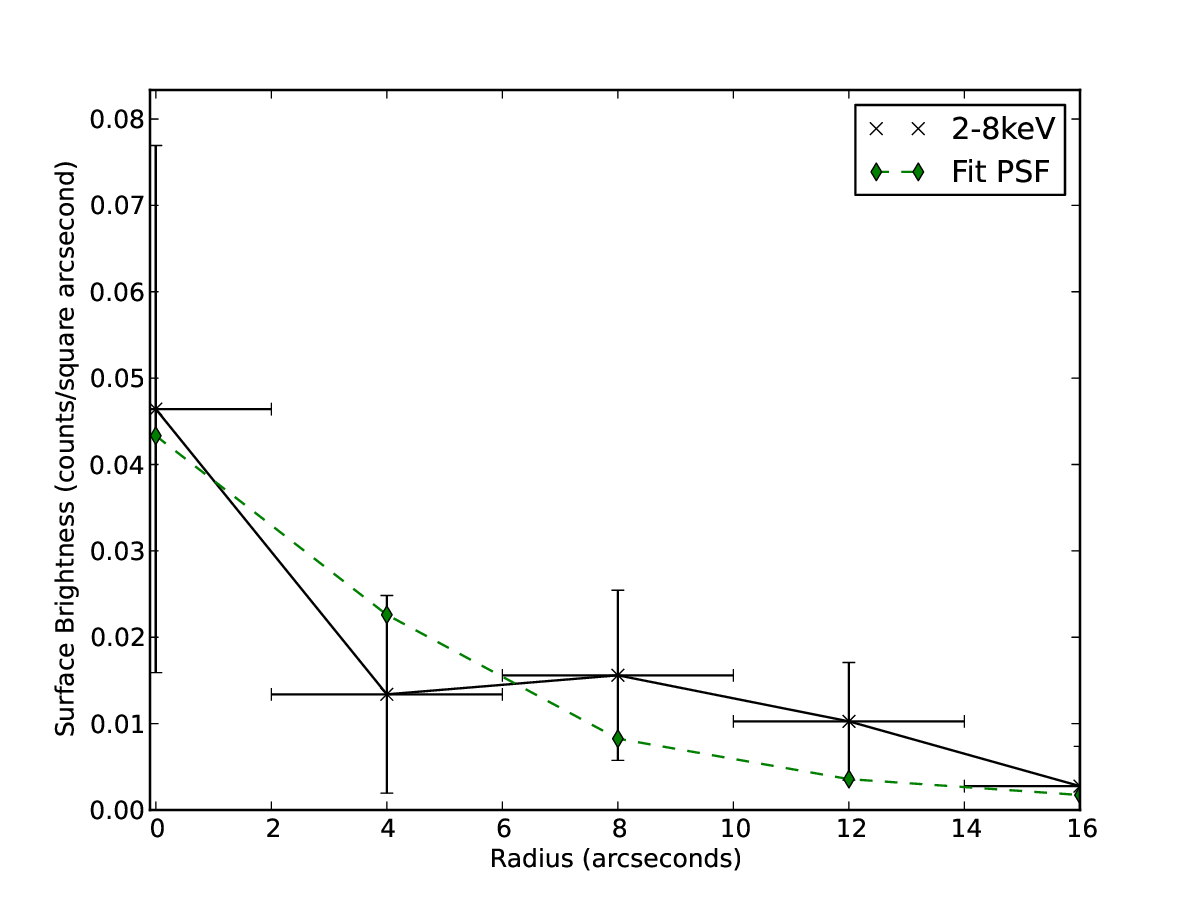}
  }					    
  \centerline{
    \includegraphics[width = 8.0cm]{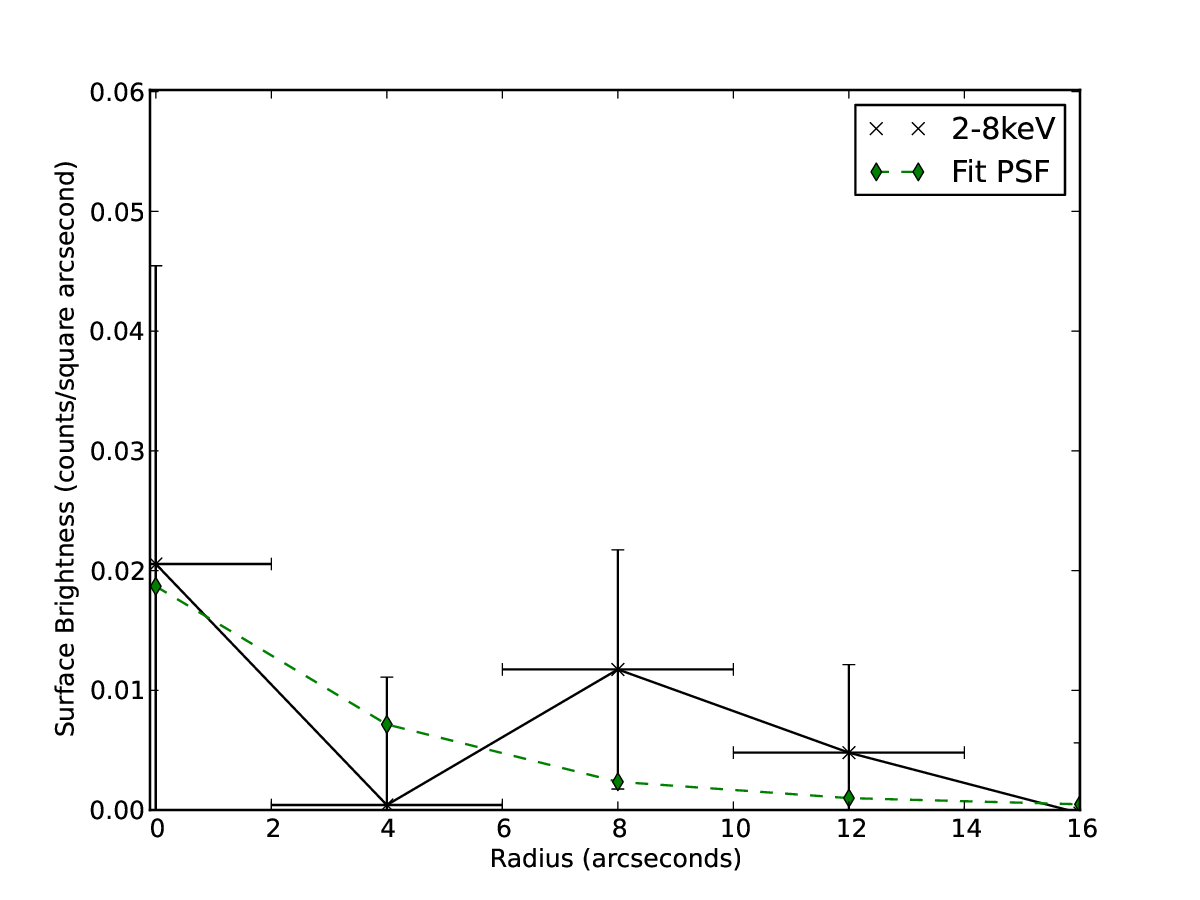}
    \includegraphics[width = 8.0cm]{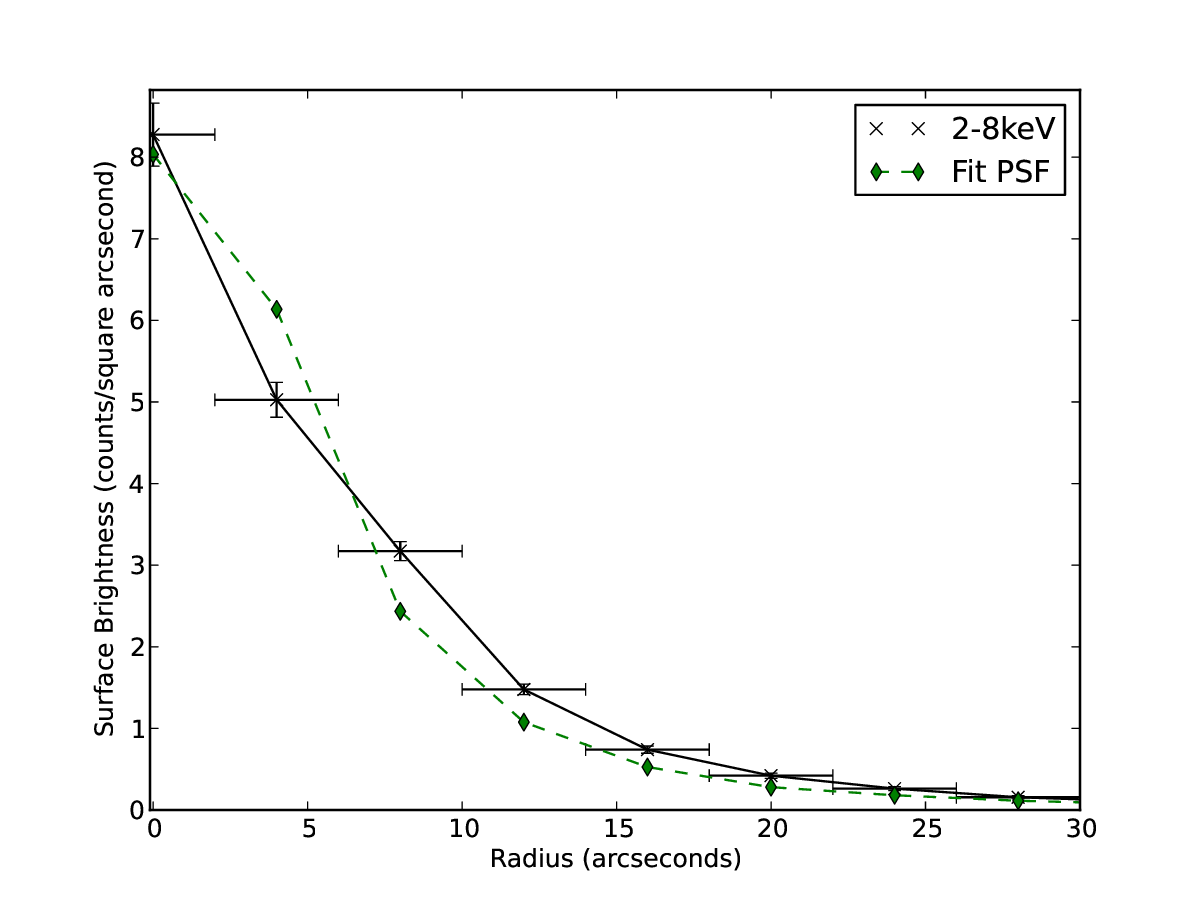}
  }                              
  \caption{Radial profile plots for our sources at 2-8 keV, as well as the models of the detector's PSF at 
5 keV at the same locations.  \emph{Top}: IRAS 18329+5950 West (18329W, \emph{Left}) and IRAS 18329+5950 East (18329E, \emph{Right}).  
\emph{Bottom}: IRAS 19354+4559 (\emph{Left}) and IRAS 20550+1656 (\emph{Right}).  A harder bandpass was chosen for this analysis as 
an AGN would be expected to be less contaminated by emission related to star formation at these energies.}
  \label{fig: RadProfile}
\end{figure*}

As hinted at above, one avenue for discerning the probable X-ray source type is to compare the radial profile of the observed flux to the PSF of the detector.  A lone AGN, 
being a point-like source, would have a profile that approximately matched the PSF's, whereas a starburst or a blend of starburst with an AGN 
would be more extended from diffuse emission.  We used the \emph{XMMSAS} task \emph{eradial} to calculate the sources' radial profile distributions 
at an energy of 2-8 keV.  We chose this energy range because an AGN which is not heavily obscured would likely provide a greater 
fractional contribution to the hard band flux than the soft band as compared to a starburst region.  The \emph{eradial} task returned 
the emission profile in concentric circular radii of each source.  It also created a model of the PSF, using the built-in ELLBETA model type, at that same location 
on the CCD chip at 5 keV, the midpoint of our hard energy range.  The ELLBETA model is based around an elliptical King profile, with an 
added Gaussian to account for spokes.  The normalization of the PSF was set by using weights inversely proportional to errors on the extracted radial profile of the data.  The background was 
fit using a region of the same area as the extraction region on the same CCD chip in order to avoid nearby sources and avoid any inter-chip variance.  Figure \ref{fig: RadProfile} shows the 
background-subtracted radial profiles.  Our systems appear to be either consistent with or 
marginally extended with respect to the PSF over our energy range.  We find the profiles of both galaxies in IRAS 18329+5950 and the single source of IRAS 20550+1656 are fully consistent 
with the PSF, while IRAS 19354+4559 may potentially be extended.  However, it is still consistent with a point source within the errors.

\subsection{Spectral Analysis}
\label{sec: specanal}
For spectral analysis, we used the entire 0.5--10 keV band.  The 90\% encircled energy radius of the \emph{EPIC PN} PSF above 2 keV is over an arcminute, so we tried to use a spectral extraction region as large as possible.  In IRAS 20550+1656, we have only one source detected and used a spectral extraction radius of $35^{\prime\prime}$, which is out to the CCD chip boundary.  In IRAS 19354+4559, again we have a single source detected.  We used a $20^{\prime\prime}$ spectral extraction radius for this system, as there are a couple of nearby sources in this field which prevented us from using a larger region.  For IRAS 18329+5950, we detected both galaxies of the system individually, which are separated by $28^{\prime\prime}$ as stated in \textsection\ref{sec: samp}.  For these sources, a spectral extraction region had to be defined carefully to avoid contamination between the two galaxies.  As such, we used a $32.^{\!\!\prime\prime}5$ radius circle instead, shown in Figure \ref{fig: Images} around 18329W.  From this region, we excised 18329E with a rectangular region centered at its coordinates, which is also shown in Figure \ref{fig: Images}.  A similar procedure was followed to create an extraction region for 18329E that had 18329W excised.  The background region was subtracted as described in \textsection \ref{sec: imanal}.  We then corrected our final luminosities based on the XMM on-axis PSF encircled energy fraction at the region size for each extracted source.

Following extraction, we performed our analysis using XSPEC v12.7.0.  We modeled each system with multiple
components in combination to account for AGN, possibly obscured, and starbursts in an attempt to recover
the observed count and energy distributions.  We were, however, limited in the complexity of our models due to low source number counts, as can
be seen in Table \ref{tab: ObservationData}.  Our fewest counts came from IRAS 19354+4559 with 46.  Both sources in IRAS 18329+5950 had around 100
counts, and IRAS 20550+1656 had over an order of magnitude greater at $\sim$3500.

\begin{figure*}
    \includegraphics[width = 6.0cm, angle=270]{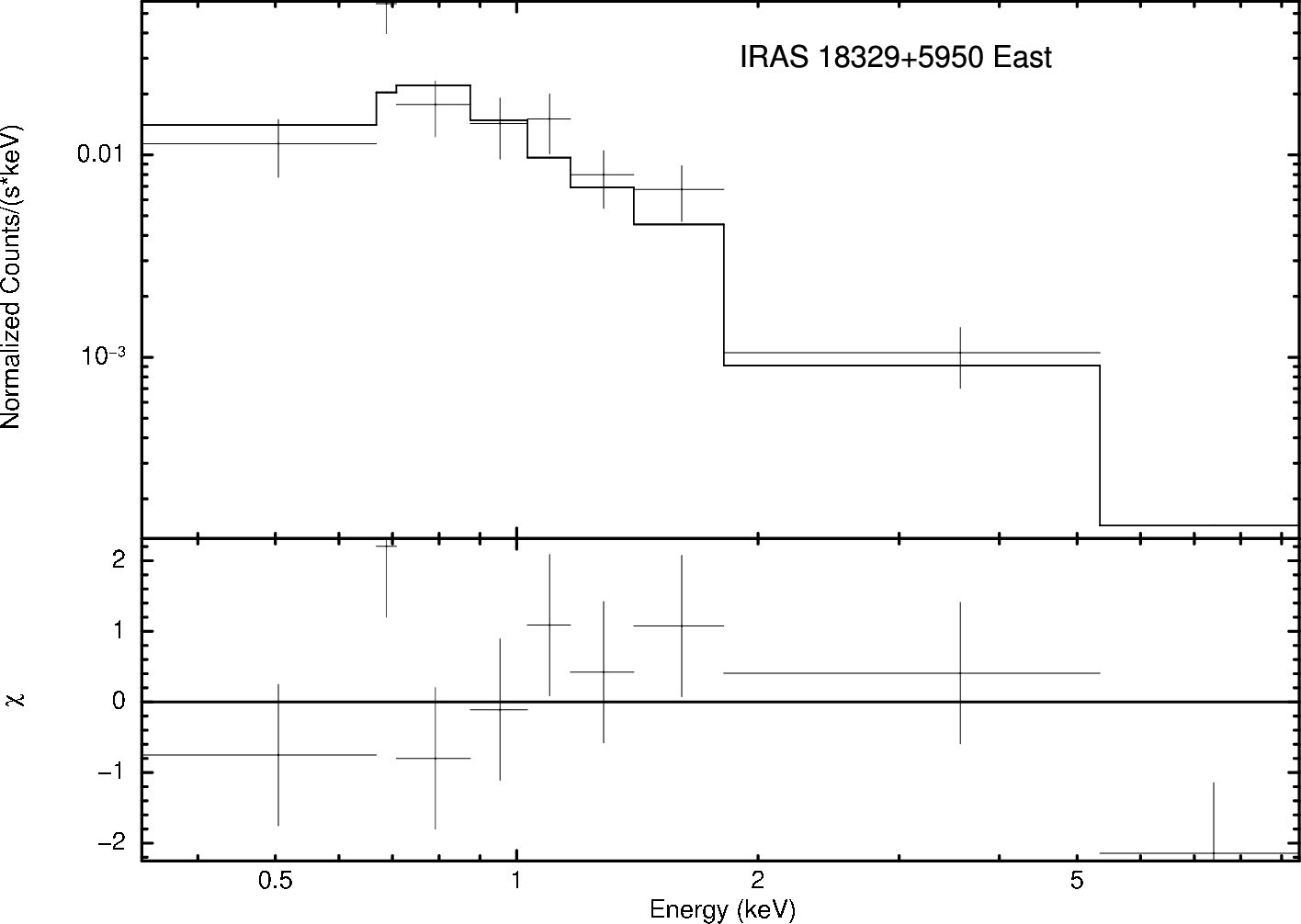}
    \includegraphics[width = 6.0cm, angle=270]{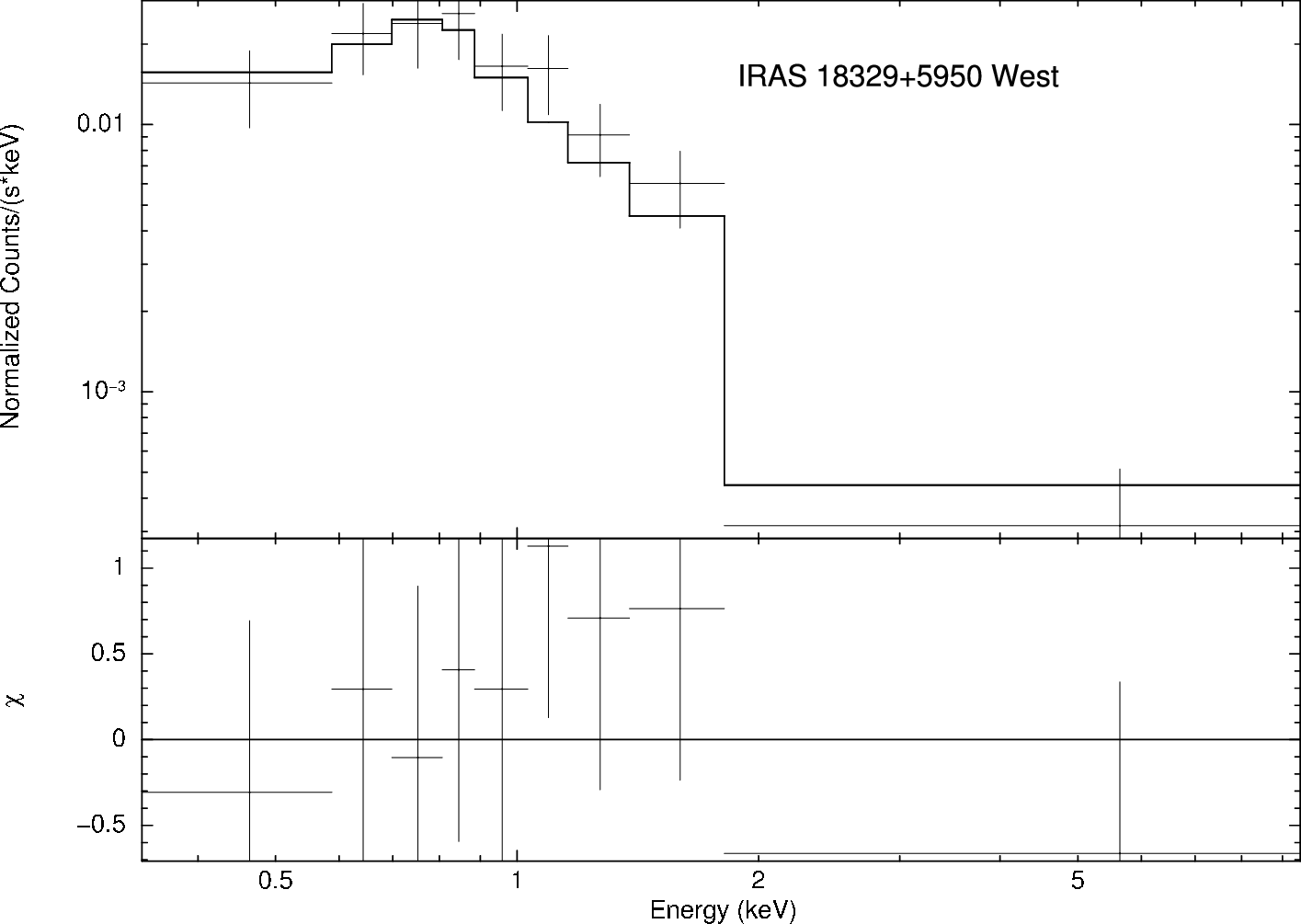}
  \caption{The best fit spectrum of the eastern source in IRAS 18329+5950 (18329E, \emph{left}) and western source (18329W, \emph{right}), both composed 
of a power law and thermal plasma component with Galactic absorption, taken from \citet{Dickey90}.  The data is rebinned for illustration purposes.  For 
comparison, the bottom panel presents the residuals between the fit and the data in units of sigma.}
  \label{fig: 6670Spec}
\end{figure*}

For IRAS 18329+5950, due to the low number counts in both sources, we decided not to bin the data and used Cash statistics rather than fit based on 
the more commonly used reduced $\chisquared$ value \citep{Cash79}.  We took into account Galactic absorption using the ``tbabs'' model \citep{Wilms00}.  
We first tried to fit the data for 18329E, shown in Figure \ref{fig: 6670Spec}, with solely an absorbed power law (C = 98.31; degrees of freedom = 113) followed by an 
absorbed power law and MEKAL \citep{Mewe85, Mewe86, Liedahl95} thermal plasma component (C = 94.60; degrees of freedom = 110), with their respective fit parameters highlighted 
in Table \ref{tab: modelComp}.  We adopt the latter model, which has a spectral index $\Gamma$ $=$ $1.8\pm0.5$ and temperature of $0.57_{-0.30}^{+0.96}$ keV.  
The component luminosities are presented in Table \ref{tab: ComponentData}; we note that this source is dominated in the hard band by the power law component.  

The system 18329W, shown in the right panel of Figure \ref{fig: 6670Spec}, was also fit by an absorbed power law (C = 98.93; degrees of freedom = 114) and then by an 
absorbed power law model and one with an additional MEKAL plasma (C = 95.70; degrees of freedom = 111), and once again a comparison between the two models is shown in Table \ref{tab: modelComp}.  
Here, the best fit value for the spectral index was slightly steeper and the temperature of the plasma slightly lower than 18329E, though they are 
consistent within their errors.  Like with 18329E, Table \ref{tab: ComponentData} indicates that 18329W derives the vast majority of its model's hard band luminosity from the 
power law and comparable amounts from both the plasma and power law in the soft band.

After correcting for the encircled energy at $32.^{\!\!\prime\prime}5$, both systems have a final soft band luminosity of $\sim10^{41}\textrm{ erg s}^{-1}$ and hard band 
luminosities of $9\times10^{40}$ and $1.4\times10^{41} \textrm{ erg s}^{-1}$ for 18329W and 18329E, respectively.  It should be noted that in both 18329E and 18329W the thermal 
plasma models were fixed at solar metallicity via the prescription in \citet{Anders89}, which is adopted by default in XSPEC.

\begin{figure}
  \includegraphics[width=5.5cm, angle=270]{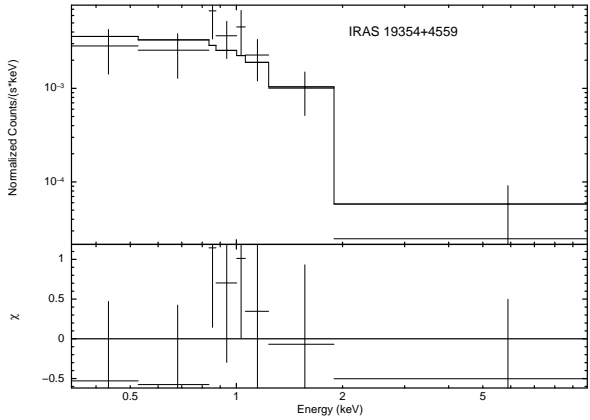}
  \caption{Similar to Figure \ref{fig: 6670Spec} but for IRAS 19354+4559.  Here the best fit model is a power law, index 2.7, with Galactic absorption.}
  \label{fig: 19354Spec}
\end{figure}

The data for IRAS 19354+4559, which can be found in Figure \ref{fig: 19354Spec}, were also fit without binning and using
Cash statistics rather than \chisquared.  Even so, multicomponent models were not well-constrained due to the paucity of counts.  As such, we tried modeling 
the system with a plasma, blackbody, and power law component separately, each with Galactic absorption.  The best fit ($C = 72.05$; degrees of freedom = 89) 
is the power law with a spectral index of $2.7_{-0.6}^{+0.6}$.  After correcting for the smaller extraction region, this system is roughly $10^{41} \textrm{ erg s}^{-1}$ in the soft band and 
$7\times10^{40}\textrm{ erg s}^{-1}$ in the hard band, as can be seen in Table \ref{tab: ComponentData}.

\begin{figure}
  \includegraphics[width = 5.5cm, angle=270]{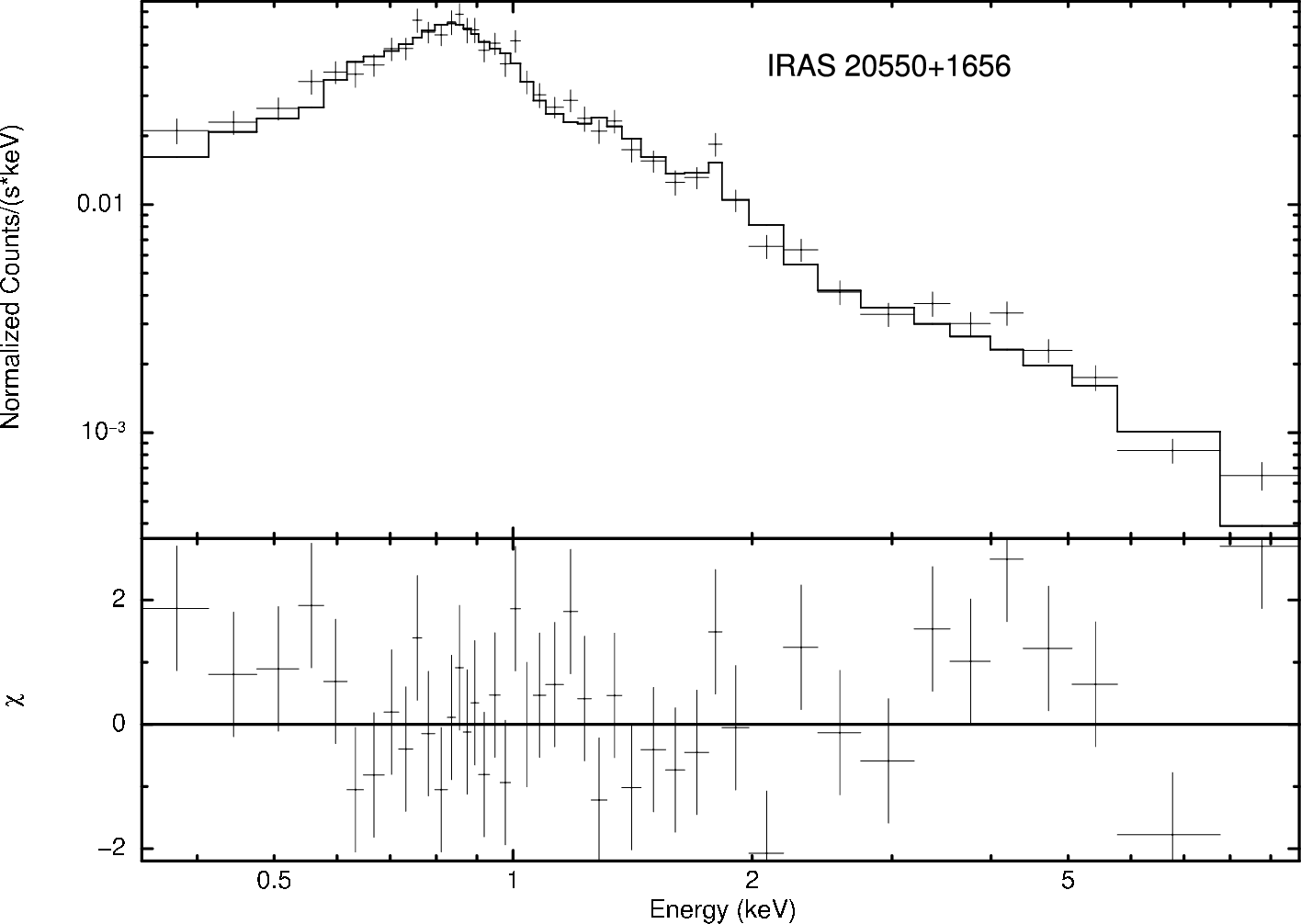}
  \caption{Similar to Figure \ref{fig: 6670Spec} but for IRAS 20550+1656.  The best-fit model has a metal-variant MEKAL plasma of temperature \emph{k}T $\sim$0.63 
keV, with the strong bump feature just below 2 keV in the spectrum fit by adopting a super-solar alpha abundance tied to silicon 
($[\textrm{Si}] = 0.5$).  The index of the power law is $\Gamma$ = 1.4.}
  \label{fig: 20550Spec}
\end{figure}

For the spectrum of IRAS 20550+1656, presented in Figure \ref{fig: 20550Spec}, we were able to bin the raw spectrum such that each bin contained a minimum of 20 counts, 
permitting us to use reduced $\chisquared$ fitting for this system.  As with all the systems, we began by fitting solely with an absorbed power law ($\chisquared$ = 360.0; degrees of 
freedom = 148).  Next, we tested an absorbed power law and MEKAL thermal plasma model ($\chisquared$ = 145.3; degrees of freedom = 145).  Comparing this to the original model, one 
can see this model is strongly favored with a reduced $\chisquared$ close to unity.  We were concerned about the bump-like features around 2 and 4.5 keV (Figure \ref{fig: 20550Spec}), however, 
and tested to see if non-solar abundances could reproduce this data.  Assessing each element individually, we found 
that the bump around 2 keV could be due to a super-solar silicon abundance.   

To investigate whether this silicon enhancement was part of a larger alpha element 
enrichment, we next fit the system with a model where all of the alpha elements were tied to the silicon value ($\chisquared$ = 132.5; degrees of freedom = 144).  
It should be noted that we also fit with both alpha elements independent from one another as well as all metals as free parameters.  Our data are unable to provide 
firm constraining power in either of these instances.  This fit, with the alpha elements fixed to the silicon value, has an abundance 
$\log\left(\frac{\textrm{Si}}{\textrm{Si}_{\odot}}\right) = [\textrm{Si}] = 0.5^{+0.1}_{-0.2}$, and is compared to the single power 
law and power law with a solar abundance plasma in Table \ref{tab: modelComp}.

We are still left with the prevalent bumps between 4-5 keV, as well as smaller ones around 1 keV.  Fitting the larger feature as a Gaussian emission line, the line has a central value 
of E = $4.3\pm0.5$ keV and line width of $\sigma$ = $0.7_{-0.5}^{+0.6}$ keV.  If this feature is an emission line from the source, it is most likely Ca XX at 3.0185A, or $\sim$4.1 keV.  
We found no reports of this line in emission in the literature, though it may have been seen in absorption in a few systems \citep{Tombesi10}.  
However, due to the tenuous detection of this line and the marginal improvement to our fit after its addition, we report the model without this component as our best fit.  
The final parameters of the fit are summarized in Table \ref{tab: modelComp}.  From Table \ref{tab: ComponentData}, 
we note that this source is dominated in the hard band by the power law component, whereas the two components are comparable in the soft band.  If we fix the plasma at solar metallicity, 
the luminosity drops by $\sim$16\% and $\sim$18\% in the soft and hard bands, respectively.  The luminosities with and without solar metallicity are within 2$\sigma$ of each other.  

How reasonable is it for us to find super-solar silicon?  We might expect to see enhanced levels of silicon from a starbursting region.  This is because starbursts are a 
rich environment of massive stars, which become Type II supernovae and can pollute the region with alpha elements.  In a recent study, \citet{Nardini13} found an elevated 
$\alpha/\textrm{Fe}$ ratio for the merging galaxy NGC 6240, known to host a dual AGN.  They argue that the heightened presence of alpha elements in NGC 6240 is consistent 
with Type II supernovae yields from \citet{Nomoto06}.  
In the Antennae Galaxies, another well-known merger, high spatial resolution revealed that the metallicity of the gas is quite variable, sub-solar in some regions and as high as 
20-30 solar in others, also arguing in favor of supernova enrichment \citep{BaldiI06, BaldiII06}.  Additionally, \citet{Araya12} report several super-solar alpha elements in their 
discovery spectra for an AGN in the bulge-less galaxy NGC 4561. 

IRAS 20550+1656 was also the subject of an extensive multiwavelength campaign, including an X-ray analysis with \emph{Chandra} as part of the GOALS project \citep{Inami10}, 
that we can compare to our data from \emph{XMM-Newton}.  They were able to distinguish two distinct objects 
in X-rays (see their Figure 5, sources labeled as ``A'' and ``C+D'') with \emph{Chandra}'s spatial resolution, whereas we see only a single source with 
\emph{XMM-Newton}.  One of their two X-ray sources was reported as extended, covering two distinct IR bright regions seen with \emph{Spitzer}.  
Using \emph{Spitzer} data on the individual sources, they concluded that source A is most likely
a starburst, obeying the \citet{Ranalli03} relations relating IR luminosity (a proxy for star formation) to X-ray luminosity (a proxy for X-ray binaries).  
For C+D they find that \citet{Ranalli03} overpredicts the X-ray values for its IR luminosity, though it should be noted that these relations were calibrated 
for less intense star forming systems.

Looking once again to this system's data from \citet{Inami10}, they report soft X-ray luminosities of $L_{\textrm{A, 0.5-2 keV}} = 6.6\: \times 10^{40} \textrm{ erg s}^{-1}$ 
and $L_{\textrm{C+D, 0.5-2\textrm{keV}}} = 1.8 \times 10^{40}\: \textrm{ erg s}^{-1}$.  Their hard X-ray (2-7keV) luminosities for 
the two sources were both $L_{\textrm{2-7\textrm{keV}}} = 10^{41}\: \textrm{ erg s}^{-1}$.  Over the entire range 0.5-7keV, then, these sum to about 50\% the luminosity we find 
for our single source over the range 2-8 keV.  One possible explanation for the differences 
in luminosity between the \emph{Chandra} and \emph{XMM} observations is that IRAS 20550+1656 is intrinsically variable, which could be suggestive of an AGN in either 
source A, source C+D, variations in the extended emission from HMXBs, or combinations thereof.  Unfortunately, with the lower spatial resolution of our observation 
we are unable to determine which of their reported sources, if only one, is responsible for our larger X-ray luminosity.

\begin{deluxetable}{llrr}

\tablewidth{0pt}
\tablecolumns{4}
\tablecaption{Final Model Component Luminosities}
\tablehead{
\colhead{IRAS Name} &
\colhead{Component} &
\colhead{$L_{\textrm{soft, 0.5-2 keV}}$} &
\colhead{$L_{\textrm{hard, 2-10 keV}}$} \\
\colhead{} &
\colhead{} &
\colhead{$10^{40}\textrm{ erg s}^{-1}$} &
\colhead{$10^{40}\textrm{ erg s}^{-1}$} 
}
  
\startdata

18329+5950E & Full Model    & $11.6_{-6.1}^{+7.5}$ &   $14.2_{-5.5}^{+5.8}$ \\ 
	    & Power Law     & $ 8.2_{-3.2}^{+3.3}$ &   $14.1_{-5.5}^{+5.7}$ \\
            & MEKAL Plasma  & $ 3.4_{-2.9}^{+4.2}$ & $0.07_{-0.06}^{+0.09}$ \\
18329+5950W & Full Model    & $ 9.0_{-4.8}^{+9.3}$ &    $8.7_{-3.3}^{+3.5}$ \\
	    & Power Law     & $ 6.4_{-2.4}^{+2.5}$ &    $8.6_{-3.3}^{+3.4}$ \\
            & MEKAL Plasma  & $ 2.7_{-2.4}^{+6.7}$ & $0.04_{-0.04}^{+0.11}$ \\
19354+4559  & Power Law     & $ 9.9_{-2.9}^{+3.1}$ &    $6.6_{-2.0}^{+2.1}$ \\
20550+1656  & Full Model    &$23.9_{-3.2}^{+3.3}$ &    $49.0_{-6.9}^{+7.2}$ \\
	    & Power Law     &$12.4_{-1.8}^{+1.8}$ &    $48.3_{-6.9}^{+7.1}$ \\
            & VMEKAL Plasma &$11.5_{-1.4}^{+1.4}$ &  $0.70_{-0.09}^{+0.09}$ \\
\enddata

\tablecomments{Comparison of X-ray luminosities of model components.  Luminosities are reported in 
$10^{40}\textrm{ erg s}^{-1}$.  Errors are calculated from the errors on the normalizations of each model component.}
\label{tab: ComponentData}
\end{deluxetable}

\begin{deluxetable*}{lrrrrr}

\tablewidth{0pt}
\tablecolumns{6}
\tablecaption{X-ray Spectral Fits}
\tablehead{
\colhead{IRAS Name} &
\colhead{Model} &
\colhead{$\Gamma$} &
\colhead{T (keV)} &
\colhead{Fit Statistic} &
\colhead{DoF}
}
   
\startdata
18329+5950 East$^{1}$ & PL 			& $2.0_{-0.3}^{+0.4}$ & \nodata		 	& 98.31 	& 113 \\ 
                      & PL + Thermal Plasma     & $1.8_{-0.5}^{+0.5}$ & $0.57_{-0.30}^{+0.96}$ 	& 94.60		& 110 \\ 
18329+5950 West$^{1}$ & PL 			& $2.1_{-0.3}^{+0.4}$ & \nodata		 	& 98.93		& 114 \\ 
                      & PL + Thermal Plasma    	& $2.0_{-0.5}^{+0.5}$ & $0.53_{-0.28}^{+2.26}$ 	& 95.70		& 111 \\ 
19354+4559$^{1}$      & PL 			& $2.7_{-0.6}^{+0.6}$ & \nodata                	& 72.05		& 89 \\
20550+1656$^{2}$      & PL 			& $1.9$		      & \nodata		 	& 360.0		& 148 \\ 
                      & PL + Thermal Plasma    	& $1.5_{-0.1}^{+0.1}$ & $0.64_{-0.04}^{+0.04}$ 	& 145.3		& 145 \\
		      & PL + Super-solar Plasma 	& $1.4_{-0.1}^{+0.1}$ & $0.63_{-0.03}^{+0.04}$ 	& 132.5		& 144 \\ 
\enddata

\tablecomments{Comparison of spectral fitting parameters for the two components of IRAS 18329+5950 and the single sources in IRAS 19354+4559 and IRAS 20550+1656.  Note that 
``DoF'' here stands for Degrees of Freedom.  $^{1}$Fit statistic is the Cash statistic.  $^{2}$Fit statistic is chi squared.}
\label{tab: modelComp}
\end{deluxetable*}

\section{Discussion}
\label{sec: disc}
With models in hand, we further investigate the likelihood of starburst and/or AGN nature of our sources. 

\subsection{Determining Star Formation Rates}
\label{sec: SFRs}
With known X-ray and IR luminosities, we now discuss the star formation rates in these systems.  There are many available prescriptions in the literature, 
and we start with the often-used \citet{Kennicutt98} IR starburst relation.  For an IR luminosity given in erg/s, 
\begin{equation}
  \textrm{SFR} \; = \;4.5\times10^{-44} L_{\textrm{IR}} \textrm{ }\msol\textrm{ yr}^{-1}\textrm{.} 
  \label{eq: KennSFR}
\end{equation}

From Equation \ref{eq: KennSFR}, we find star formation rates (SFRs) in the range of 75-150 $\msol\textrm{ yr}^{-1}$ for our three systems, with IRAS 20550+1656 being the strongest.  
We have listed the star formation and IR luminosities for each individual system in Table \ref{tab: SFRLuminTable}. 
As is typically expected of mergers and LIRGs in general, these are relatively large SFRs, implying that at least one of the galaxies in each pair 
is likely in a starburst phase.  The underlying assumption in this and other relations is that all of the UV emission from O and B stars, whose presence is indicative of 
recent star formation, is absorbed by dust surrounding the star forming region and re-radiated into the infrared.  Not all of this UV radiation is absorbed, however.  
For example, \citet{MirallesCaballero12} observed some star clusters with as much as 15\% unobscured UV radiation.  As such, we decided to also try a more recent 
SFR relationship that takes both IR and UV radiation into account \citep{IglesiasParamo04, IglesiasParamo06, Hirashita03, Bell03}:
\begin{equation}
 \sfrAll,
 \label{eq: ipSFRtotal}
\end{equation}
where
\begin{eqnarray}
 \sfrIR,
 \label{eq: ipSFRir} \\ 
 \sfrUV.
 \label{eq: ipSFRuv} 
\end{eqnarray}
and all luminosities here are taken in erg/s.

The $\eta$ factor in Equation \ref{eq: ipSFRtotal} is a correction factor (between zero and unity) for the fraction of IR emission that is cirrus in nature rather 
than directly related to recent star formation.  The value of $\eta$ for a specific galaxy is difficult to ascertain.  \citet{Bell03} find that $\eta \sim 0.09$ for galaxies 
with $\log(L_{\textrm{IR}}/\lsol)$ > 11, whereas galaxies below this threshold have $\eta \sim 0.3$.  As all of our systems are in the former 
regime, we adopt a value of $0.09$ in all of our analyses when necessary.  For the NUV luminosities, we used GALEX data presented in \citet{Howell10} 
for two of our systems.  The NUV term in Equation \ref{eq: ipSFRtotal}, however, has less than a 1\% effect on the total SFR for IRAS 18329+5950 and 
IRAS 20550+1656, so we proceed using only the IR SFR, Equation \ref{eq: ipSFRir} modified by $\eta$, for our systems.  We caution that IRAS 19354+4559 has the lowest IR luminosity, 
and thus may have a larger relative UV contribution than the other systems for which GALEX data was available.  Overall, the SFRs obtained in this way are slightly below those from 
the \citet{Kennicutt98} formalism, ranging from 72-140 $\msol\textrm{ yr}^{-1}$.  These are also listed in Table \ref{tab: SFRLuminTable} for the individual 
systems in our sample for comparison.

\citet{Inami10} find a SFR of $120$ $\msol\textrm{ yr}^{-1}$ for the non-nuclear IR source (labeled ``D'' in their paper) in IRAS 20550+1656.  
Comparing to our value for the system as a whole, this would imply that most of the star formation in the system is happening in this non-nuclear region.

\subsection{X-ray Origins}
\label{sec: X_IR}
Our next objective was to determine the major contributor to our systems' X-ray luminosities.  Are they related to star formation, AGN activity, or both?

To accomplish this, we compare our luminosities to literature predictions for X-rays associated with star formation.  For this task, 
we employ the relationship of \citet{Mineo14}:
\begin{equation}
 \MineoXray
 \label{eq: MineoX}.
\end{equation}
For consistency, we use star formation rates from \citet{IglesiasParamo06}, here simplified to Equation \ref{eq: ipSFRir} for our IR luminous systems, rather 
than \citet{Kennicutt98}, as these were the rates used to calibrate Equation \ref{eq: MineoX}.  Using the latter, however, results in a 6\% increase in the predictions.      

Since these relationships pertain directly to starbursts via their diffuse emission and integrated X-ray binary luminosity, if they are similar to the luminosities 
we find from our observations, we may infer that our systems, at least in the X-ray range, are most likely dominated by star formation rather than AGN.  If a strong, unabsorbed 
AGN is present, however, we would expect a notable excess in X-ray luminosity compared to these predictions.  

\begin{figure}
  \includegraphics[width=9.5cm]{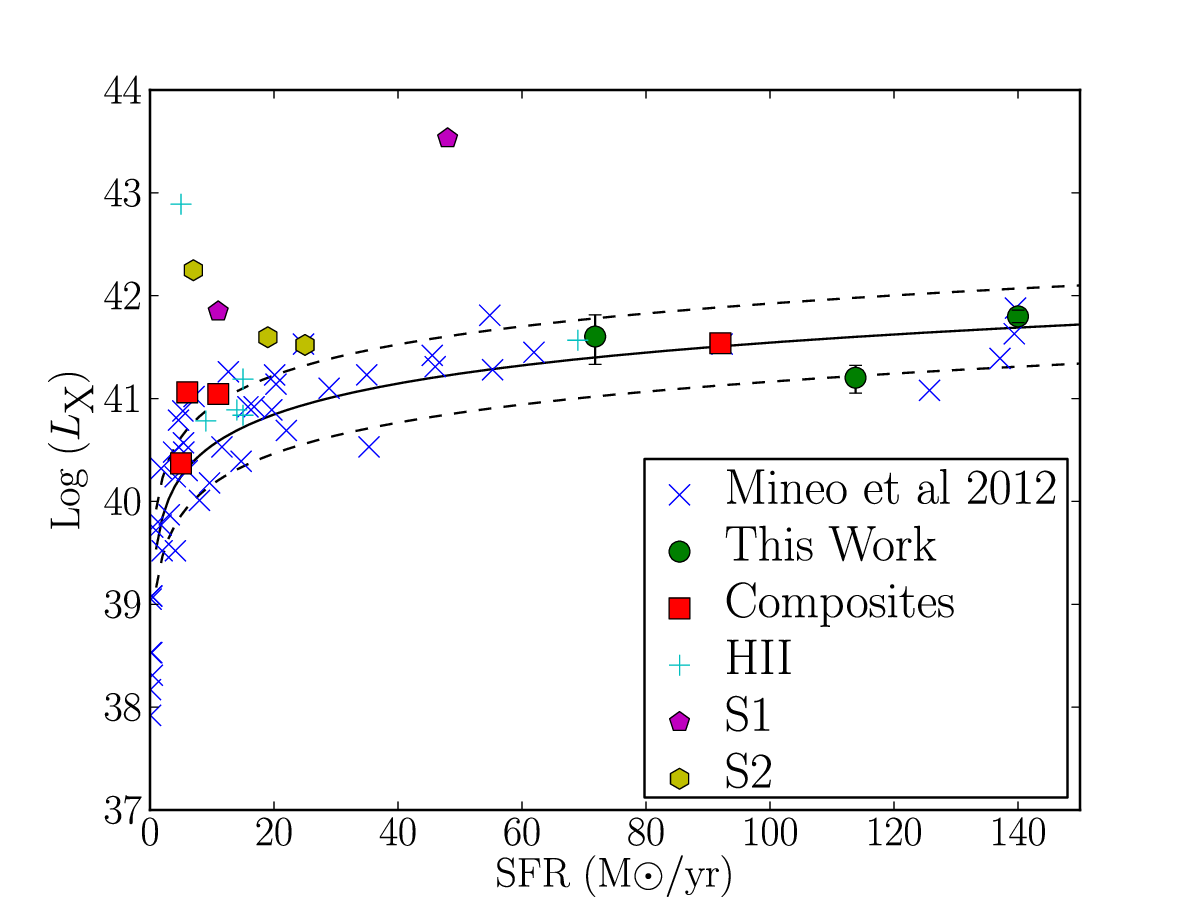}
  \caption{Comparison between the modeled X-ray luminosities of our three systems with predictions from Equation \ref{eq: MineoX} (solid 
line).  The dashed lines are the $1\sigma$ errors on this relation.  The data for composites, HII systems, Seyfert 1s, and 
Seyfert 2s were compiled by \citet{PereiraSantaella11}, and taken from \citet{Lehmer10} for NGC 23, NGC 5653, and Zw049.057; \citet{Miniutti07} for 
MCG-03-34-064; \citet{Levenson05} for NGC 7130; and \citet{Blustin03} for NGC 7469.  Our LIRGs tend to lie in a region occupied by HII systems, composite 
AGN/starbursts, and heavy star-forming galaxies.  It should be noted that these galaxies' luminosities are from 0.5-10 keV, whereas Equation \ref{eq: MineoX} 
is defined for 0.5-8 keV by \citet{Mineo14}, and are thus systematically offset from the relation apart from their scatter by a small amount ($\leq$0.1dex).  
The systems presented in this paper and from \citet{Mineo12I} are on a 0.5-8 keV scale.  In order of increasing star formation, our points are for 
IRAS 18329+5950, IRAS 19354+4559, and IRAS 20550+1656.}
  \label{fig: MineoPlots}
\end{figure}

In Figure \ref{fig: MineoPlots}, we plot our derived luminosities against star formation rate and include other known AGN and AGN-starburst composite systems.  
From this, it is evident that the best-fit models for each of our systems described in \textsection\ref{sec: specanal} give 0.5-8 keV luminosities that are consistent with the 
expected values from their SFRs.  IRAS 19354+4559 is on the lower end of this range, but we remind the reader that this particular system had the least constrained model 
as discussed in \textsection\ref{sec: specanal}.  The values of these luminosities, both predicted and modeled, are shown for convenience in Table \ref{tab: SFRLuminTable}.  
If we adopt our solar metallicity luminosity for IRAS 20550+1656, it is closer to the predicted value from star formation, but both are consistent.  In their 
discussion, \citet{Mineo14} compare the calibration of their relationship to others found in the literature. They find that there are two primary differences in how 
various studies have generated these associations.  The first is how each work proxies the star formation rate of the galaxies studied, and the second is their adopted model 
for X-ray binary emission.  Combined, these two issues can drop the normalization of Equation \ref{eq: MineoX} by a factor of 2 or raise it by a factor of 1.5.  We elect to work 
with Equation \ref{eq: MineoX} because of the care in handling both the diffuse and source (here, X-ray binary) emission related to star forming regions.

\begin{deluxetable*}{lrrrrr}

\tablewidth{0pt}
\tablecolumns{6}
\tablecaption{IR and X-ray Luminosities}
\tablehead{
\colhead{IRAS Name} &
\colhead{$L_{\textrm{IR}}$} &
\colhead{$\textrm{SFR}_{\textrm{K}}$} &
\colhead{$\textrm{SFR}_{\textrm{IP}}$} &
\colhead{$L_{\textrm{0.5-8 keV}} \textrm{(Pred)}$} &
\colhead{$L_{\textrm{0.5-8 keV}} \textrm{(Obs)}$} \\
\colhead{} &
\colhead{$(10^{44}\textrm{ erg/s})$} &
\colhead{$(\msol\textrm{ yr}^{-1})$} &
\colhead{$(\msol\textrm{ yr}^{-1})$} &
\colhead{$(10^{40}\textrm{ erg/s})$} &
\colhead{$(10^{40}\textrm{ erg/s})$} 
}
  
\startdata

18329+5950 & 17 & 76  &  72 &  $25^{+35}_{-15}$ & $ 40^{+25}_{-19}$ \\ 
19354+4559 & 27 & 121 & 114 &  $40^{+56}_{-23}$ & $ 16^{+5}_{-5}$ \\ 
20550+1656 & 33 & 149 & 140 &  $49^{+69}_{-29}$ & $ 63^{+9}_{-9}$ 

\enddata

\tablecomments{Star formation rates, predicted luminosities, and modeled luminosities related to our systems.  The IR fluxes are calculated 
from Equation \ref{eq: IRflux} using the specific flux values presented in the RBGS \citep{Sanders03}. 
The star formation rates are calculated from \citet{Kennicutt98} and \citet{IglesiasParamo06}, reproduced 
in-text as Equation \ref{eq: KennSFR} and Equation \ref{eq: ipSFRir}, respectively.  The predicted X-ray luminosities from SFR alone are found using 
$\textrm{SFR}_{\textrm{IP}}$ values and Equation \ref{eq: MineoX}, taken from \citet{Mineo14}.  Our X-ray values are from our best fit spectral 
models described in the text.}
\label{tab: SFRLuminTable}
\end{deluxetable*}

\subsection{AGN in LIRGs?}
\label{subsec: resumen}
We have employed several methods to determine whether or not our sample contains AGN or dual AGN.  One indicator 
would be X-ray point sources.  From Figure \ref{fig: Images} we see that all three systems
are detected in the entire XMM energy bandpass and that IRAS 18329+5950 is composed of two distinct sources.  If these X-ray sources were due solely
to the presence of an AGN, one would expect that their radial profiles would be similar to the detector's PSF.  From Figure
\ref{fig: RadProfile} we see that, for the most part, the 2-8 keV emission is consistent within errors of the modeled PSF at 5 keV.  Given that the 
FWHM for the PN detector on \emph{XMM-Newton} is about an 12.$^{\!\!\prime\prime}5$, amounting to $\sim$8 kpc in the closest system (IRAS 18329+5950), 
it is impossible to say definitively that the X-ray source is solely from point-like emission. 

The second test was to compare the spectra of each system to starburst and AGN templates.  Both sources in IRAS 18329+5950 and 
the single source in IRAS 20550+1656 were composed of a MEKAL plasma, typically associated with ionizing photons from OB stars and their 
supernovae, as well as a power law which could be from either an AGN or X-ray binaries.  As for their spectral indices, X-ray binaries are expected, 
from \citet{Persic02}, to have power law slopes of $\sim$1.2, though as stated in \textsection\ref{sec: intro}, the full range is $\Gamma\sim$ 1-2.4, 
typically between 1-2.  The power law components of both systems of IRAS 18329+5950 as well as IRAS 20550+1656 are within this range.  
IRAS 19354+4559 is steeper than the others, but all of our systems are consistent within error bars with both HMXBs and AGN spectral indices.  

Lastly, we investigated how our systems' X-ray luminosities compared to their star formation rates.  All of our systems' 
X-ray luminosities lie within 1$\sigma$ of the prediction from star formation alone.  This implies that 
their X-ray output is in a regime where it can be attributed mostly, if not entirely, to processes tied to star formation, such as XRBs and diffuse emission.

In all of these tests, we do not find robust evidence for AGN activity in any of our sources, though we cannot definitively rule out AGN activity either.

This, then, raises the question:  Why there are no dual AGN, or even single AGN, signatures in our systems?  Ideally, obtaining more counts for each 
our systems would allow us to more conclusively state whether there are no AGNs or simply weaker ones.  Additional counts would permit us to fit more complex 
models with higher constraining power that could better discriminate between starburst and AGN templates.  This is a severe problem with IRAS 19354+4559 
in particular, as we could only fit it with a single power law component which clearly has a soft excess, as shown in Figure \ref{fig: 19354Spec}.  
Additional spatial resolution, such as with \emph{Chandra}, would be beneficial as well as it could demonstrate more clearly the extension of the X-ray emission, 
as in the case of IRAS 20550+1656 in \citet{Inami10}.  The biggest obstacle among all our data was perhaps the loss of large portions of our exposures due to background, 
especially in IRAS 18329+5950 and IRAS 19354+4559.  

\begin{figure}
  \includegraphics[width = 9.5cm]{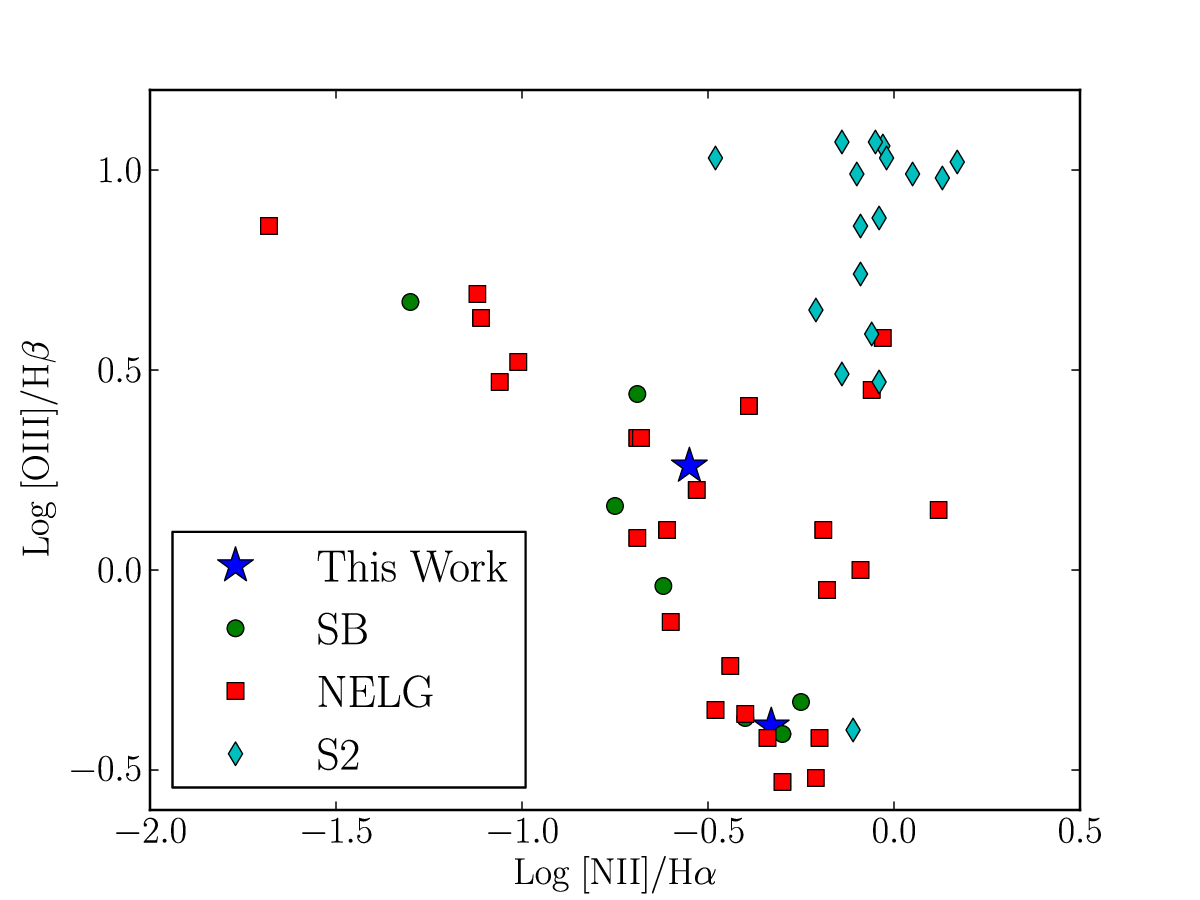}
  \caption{BPT diagram from \citet{Veilleux87} with IRAS 18329+5950 (topmost star) and IRAS 20550+1656 (bottommost star) added.  They occupy the same parameter space as narrow emission line galaxies 
and starburst systems.}
  \label{fig: BPT_Vostebrock}
\end{figure}

It is possible that any AGN present in our systems is Compton thick with large ($>10^{24}$ $\textrm{cm}^{-2}$) column densities, hiding nuclear 
activity.  Knowing that these systems are heavily obscured in the optical, we investigated the location of our systems on a BPT diagram \citep{Baldwin81} of N[II]/H$\alpha$ 
against O[III]/H$\beta$.  In Figure \ref{fig: BPT_Vostebrock} we show IRAS 18329+5950 and IRAS 20550+1656 (H$\beta$: \citealp{Kennicutt09}, [NII], [OIII], and H$\alpha$: \citealp{Moustakas06}) 
alongside data and classifications taken from \citet{Veilleux87} as reference, and see that they lie in the same region as the starbursts (SBs) and narrow emission line galaxies (NELGs, those 
galaxies with profiles akin to HII and LINER systems), similar to Figure \ref{fig: MineoPlots}.  We also looked for 
mid-IR lines in the literature that would be less subject to internal extinction than the optical ones.  The only such data found was for IRAS 20550+1656, where a combination of 
PAH, [NeII], and [NeIII] emission suggested the system was a starburst rather than an AGN \citep{Inami10}.      

Our data are consistent with there being no prominent AGN present.  What does this imply?  The simulations from \citet{Hopkins06} have a large fraction of SMBH growth due to 
galaxy mergers and imply that the merger rate should map relatively well onto the quasar activation rate.  They find that major accretion occurs even during the first passage.  
The peak quasar activity is associated with the final merger after dust blowout, though lower luminosity activity is predicted both before and after this phase.  However, 
it is possible that Class III mergers may be too early in the merger process to have strong AGN activity (i.e. enough to distinctly surmount starburst 
emission) in the galaxies involved.  

Unfortunately, it is observationally ambiguous how long any single merger has been going on, and thus merging systems without AGN provide little hint as to how far ``behind schedule'' 
they are with respect to simulations such these. This most luminous phase of the AGN is the likely culprit of the correlations presented in \textsection\ref{sec: intro}.  Other studies, however, argue 
that fueling during mergers is \emph{not} the main avenue for activating a central SMBH.  \citet{Kocevski12} find in a sample of 72 systems at larger redshifts (1.5 < z < 2.5) that AGNs are 
no more likely to possess a disturbed morphology than their control sample systems.  In a more local sample (0.3 < z < 1) of 140 AGN with \emph{XMM-Newton} data, \citet{Cisternas11} find that 
85\% show no signs of recent major mergers, which is not significantly different (<1$\sigma$) from this same fraction of their control sample.  \citet{Grier11}, using the Spitzer Infrared 
Nearby Galaxies Survey, find that 60\% of galaxies host AGN from X-ray data and these galaxies are not in merging systems.  It is possible, however, that most of their black hole 
growth happened in previous major mergers.  The presence of AGN in galaxies with pseudobulges (\citealp{Mathur12} and references therein) clearly points to an alternative path of black hole growth.

If the SMBHs in our systems are actually not accreting, rather than simply too weak to be discerned from star formation, then we must look at simulations with more caution.  Does intense 
nuclear star formation or some other merger-related process prevent the ample supply of gas and dust from fueling the SMBHs at early stages?  Given the small number of systems reported 
with dual AGN and the larger number of attempts to find them through varied methods, there could be an issue with this phase of the hierarchical growth route altogether.  
Some (possibly substantial) fraction of SMBHs may not accrete much material during mergers, in which case one might expect most of the AGN activity to occur during slower 
secular accretion processes (e.g., \citealp{Mathur12, Kocevski12}).  Or, maybe the lifetime of dual AGN activity is extremely short and thus hard to observe, as some studies have suggested
\citep{VanWassenhove12}.  The star formation processes also are quite powerful in mergers, and can possibly dwarf lower luminosity AGN activity that occurs as the 
SMBHs begin to accrete matter, as can be seen in Figure \ref{fig: MineoPlots}.  Any one of these would make finding binary AGN in mergers quite difficult.  Unfortunately, our data 
cannot discriminate between these scenarios and we are unable to definitively state which, if any, is the true culprit at this time.  

\section{Conclusion}
\label{sec: conclusion}
We present the results from a search for dual AGN.  
Based on the environment arguments presented in \textsection\ref{sec: intro}, we selected nearby LIRGs 
(tending to have lower $L_{\textrm{IR}}$ than those observed at higher redshifts) with galaxy separations near the 
resolution limit of the \emph{XMM-Newton} and with evidence of interaction.  From our imaging analysis, the only system 
in which two distinct X-ray sources are resolved is IRAS 18329+5950.  Only one of these, the eastern source in IRAS 19354+4559, 
is possibly inconsistent with the PSF, and the rest of our systems seem to be dominated by emission from the central few kiloparsecs.

The X-ray luminosities of IRAS 18329+5950, IRAS 19354+4559, and IRAS 20550+1656 are 
all within $1\sigma$ of the predicted value from the \citet{Mineo14} relationships mapping star formation rate to X-ray luminosity.  
This suggests that the X-rays for these galaxies arise from star formation rather than AGN.

The data for each of our systems is not of sufficient quality to find, nor conclusively rule out, the presence of AGN.  
These results could be improved upon by searching for other nearby LIRG systems with resolvable nuclear regions.  If such systems 
continue, as the three presented in this paper, to be consistent with star formation related emission alone, then we will 
be able to place observational constraints on the predictions from simulations presented in \textsection\ref{sec: intro}.

\acknowledgements{}
Based on observations made with the NASA/ESA Hubble Space Telescope, and obtained from the Hubble Legacy Archive, which is a collaboration between the Space Telescope Science Institute (STScI/NASA), the Space Telescope European Coordinating Facility (ST-ECF/ESA) and the Canadian Astronomy Data Centre (CADC/NRC/CSA). 

We thank the XMM Science Team for their invaluable advice with data processing and useful assistance, those who worked on the Digitized Sky Survey, and those who 
maintain their online services for obtaining the optical image of IRAS 19354+4559.  D.M. is also extremely grateful to Anjali Gupta, Ben Shappee, Claudia Araya Salvo, Joe Antognini, 
Garrett Somers, Scott Adams, Michael Fausnaugh, and Obright Lorain for help along the way.  We are also grateful to the anonymous referee for exceptionally useful comments.

\bibliographystyle{apj}
\bibliography{references2}

\end{document}